\documentclass[preprint,showpacs,12pt,nofootinbib]{revtex4}
\usepackage{bm} 
\usepackage{amsmath}     
\usepackage{epsfig}
\usepackage{graphicx}
\usepackage{pdfpages}
\usepackage{color}   
\usepackage{graphicx}

\begin{document}

%\title{Response functions and correlations : Particle physics experiments and correlated nucleons}

\title{Multinucleon excitations in neutrino-nucleus scattering: connecting different microscopic models for the correlations}

\author{G. Chanfray}
\affiliation{Univ Lyon, Univ Claude Bernard Lyon 1, CNRS/IN2P3, IP2I Lyon, UMR 5822, F-69622, Villeurbanne, France}

\author{M. Ericson}
\affiliation{Univ Lyon, Univ Claude Bernard Lyon 1, CNRS/IN2P3, IP2I Lyon, UMR 5822, F-69622, Villeurbanne, France}
\affiliation{Theory department, CERN, CH-12111 Geneva, Switzerland}

\author{M. Martini}
\affiliation{IPSA-DRII,  63 boulevard de Brandebourg, 94200 Ivry-sur-Seine, France}
\affiliation{Sorbonne Universit\'e, Universit\'e Paris Diderot, CNRS/IN2P3, Laboratoire
de Physique Nucl\'eaire et de Hautes Energies (LPNHE), Paris, France}

\begin{abstract}
%\abstract{
%Since the MiniBooNE measurement of the neutrino charged-current quasielastic-like cross section on carbon and the proposed explanation of its behavior based on the inclusion of multinucleon ejection channel, 
The problem of nucleon-nucleon correlations and  meson exchange currents has been vividly debated in  connection with the neutrino-nucleus cross sections. 
In this work we focus on nucleon-nucleon correlations by discussing a formal correspondence between the approaches based on independent particles and the \textit{ab initio} approaches involving correlated wave functions. 
We use a general technique based on unitary transformation mapping the Fermion operators relative to bare nucleons into quasi-particle operators relative to dressed nucleons. We derive formulas for spectral functions, response functions, momentum distribution, separation energy, general enough to be applied with any kind of effective nucleon-nucleon interaction. 
We establish the relation between the  non-energy-weighted sum rule and the Fermi sea depopulation. 
With our tools we evaluate whether approaches based on effective interactions are compatible with the expected amount of correlations coming from \textit{ab initio} calculations. For this purpose we use as a test the Fermi sea depopulation and the value of the kinetic energy per nucleon.
\end{abstract}

\maketitle
%%%%%%%%%%%%%%%.%%%%%%%%%%%%%%%%%%%%%%%%%%%%%%%%%%%%%%%%%%%%%%%%%%%%%%%%%%%%%%�
%\MM{28/06/2021 VERSION FINALE POUR UNE DERNIERE RELECTURE}
\section{Introduction}\label{Intro}
%%%%%%%%%%%%%%%%%%%%%%%%%%%%%%%%%%%%%%%%%%%%%%%%%%%%%%%%% is t%%%%%%%%%%%%%%%%%%%%%%%%

In neutrino experiments in the GeV region, the range of atmospheric and accelerator-based neutrino experiments, 
as neutrinos interact weakly,
%, in order to increase the  number of interactions 
nuclei are used as detectors in place of nucleons. 
In the first analysis the nucleus was seen as a collection of independent nucleons and  the neutrinos were assumed to interact with individual nucleons. This follows from the independent particle picture of the nucleus seen as an assembly of nucleons, bound by a smooth mean field potential. 
%Their Fermi motion is taken into account in the analysis of the experiments.
This is indeed the important foundation of nuclear physics. As Bethe emphasized,  "The most striking features of finite nuclei is the validity of the shell model. Nuclei can be very well described by assigning quantum numbers to individual nucleons" \cite{Bethe}. This concept is implemented in the Hartree-Fock scheme where the nuclear ground state is described by a Slater determinant where only certain levels, with eigenenergy calculated with an effective interaction, are occupied \cite{Vautherin}.

%For the nucleons of a free Fermi gas, their response  to an external probe is limited to the energy $\omega$ and momentum ${\bf{q}}$ relation, $\omega= Q^2/(2M)$ where $Q^2={\bf{q}}^2 -  \omega^2$ and $M$ is the nucleon mass, with eventual correction due to the Pauli blocking and Fermi motion effects. Now in reality nucleons of the nucleus are not free, they are correlated with the neighboring nucleons. The issue is then to find how the simple energy momentum relation of the free nucleon is modified by these interactions. This is not an easy problem since it involves all the complexities of the nuclear interactions. The simple line of response in the  $(\omega, |{\bf{q}}|)$ plane becomes a whole region. In fact the necessity of their introduction has not always been immediately realized. This was the case for neutrino interactions, as illustrated in the following.

The first realization that the neutrino-nucleus cross section  is not well modeled in this way goes back to the years 2009/2010 when the 
MiniBooNE measurement \cite{Katori2009,MiniBooNe2010} of the charged-current quasielastic-like  (CCQE-like) cross section turned out to be in strong disagreement with the predictions of the Fermi gas model (the model then implemented in all the Monte Carlos). The evaluations based on the assumption of interactions between neutrinos and independent nucleons strongly underestimate the MiniBooNE cross section. The explanation of this disagreement came from the Lyon group \cite{MECM2009}: nucleons are correlated, via short-range correlations (SRC) and meson exchange currents (MEC), which implies the possible ejection of a pair of nucleons. This process, which increases the cross section, is referred as 2 particle-2 hole (2p-2h), or n particle-n hole (np-nh).
After this suggestion, the interest on the multinucleon emission channel rapidly increased. Different kinds of calculations have then been performed. They can be divided in three categories. The first one starts from an independent particle model (IPM) and on top of it calculates the 2p-2h contributions to the neutrino-nucleus cross section. 
This is the case for the approaches of the Lyon \cite{MECM2009,MECM2010,MEC2011,ME2013,ME2014,EM2015,Martini:2016eec} and Valencia 
\cite{Nieves2011,Nieves2012,Nieves2013,Gran:2013kda,Sobczyk:2020dkn,Bourguille:2020bvw} groups, which start from a local Fermi Gas, and of the Ghent \cite{Pandey:2014tza,VanCuyck:2016fab,VanCuyck:2017wfn} and La Plata \cite{Martinez:2021gex} groups, which 
start respectively from non-relativistic (Skyrme) and relativistic (Walecka) mean field approaches\footnote{We remind that in the case of Lyon, Valencia and Ghent groups long range nuclear correlations are also taken into account, via the Random Phase Approximation (RPA).}.
In all these calculations, the 2p-2h excitations are evaluated by generalizing microscopic models 
%introducing explicitly SRC and MEC contributions involving one-pion exchange and $\Delta$ excitation, which were
previously developed to study electron scattering, pion and photon absorption \cite{AEM84,Oset:1987re,Gil:1997bm,Ryckebusch:1993tf,Ryckebusch:1997gn}, hence constrained by these processes. 
In the second category, such as Green's function Monte Carlo  \cite{Lovat2014,Lovat2015,Lovato:2017cux,Lovato:2020kba} and spectral function  \cite{Benhar:2005dj,Benhar2016,Rocco:2015cil,Rocco:2018mwt,Barbieri:2019ual} approaches, one starts from a correlated wave function adding in some cases MEC contributions. In these approaches the 2p-2h excitations arising from SRC are automatically incorporated from the beginning. 
The third category is represented by models directly constrained by electron scattering phenomenology. This is the case of the superscaling (SuSA) approach and of GiBUU. The superscaling (SuSA) approach \cite{Amaro:2004bs}, as well as its updated version, called SuSAv2 \cite{Gonzalez-Jimenez:2014eqa}, 
is a phenomenological model which provides by construction a good description of inclusive electron scattering data in the quasielastic region. The multinucleon excitations are included via a microscopic fully relativistic calculation of 2p-2h excitations, induced by MEC, of the Fermi gas \cite{Simo:2014wka,Simo:2016ikv,Megias:2016fjk,RuizSimo:2016ikw,RuizSimo:2017onb,Amaro:2017eah,RuizSimo:2017hlc,Megias:2017cuh}. In the case of GiBUU instead, microscopic ingredients, such as a mean field potential, are taken into account to describe the 1p-1h excitations \cite{Leitner:2008ue} while 2p-2h excitations are included via an empirical spin-isospin response deduced from electron scattering data \cite{Gallmeister:2016dnq}.

\begin{figure}
 \centering
  \includegraphics[width=0.6\textwidth]{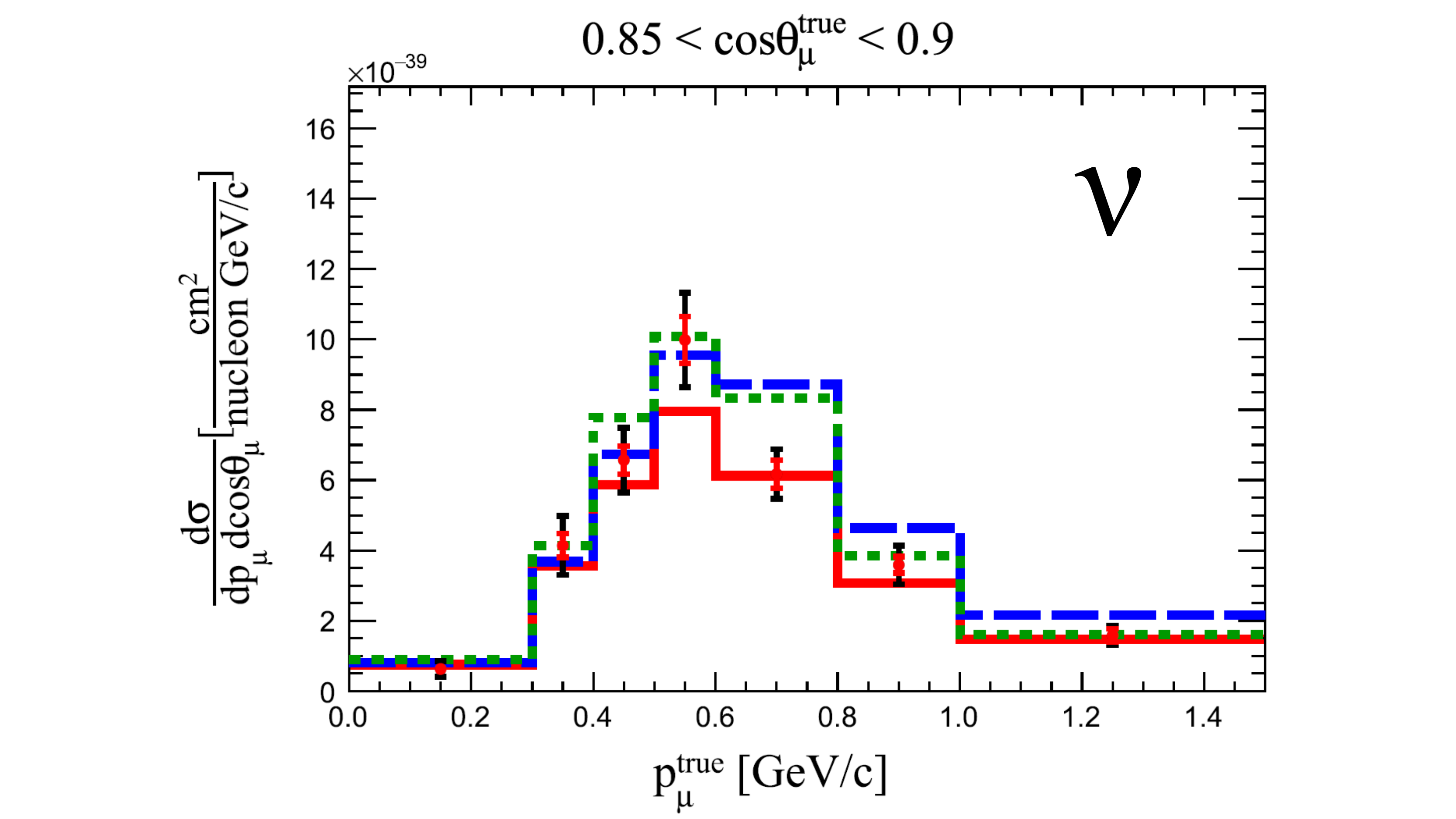}
  \includegraphics[width=0.6\textwidth]{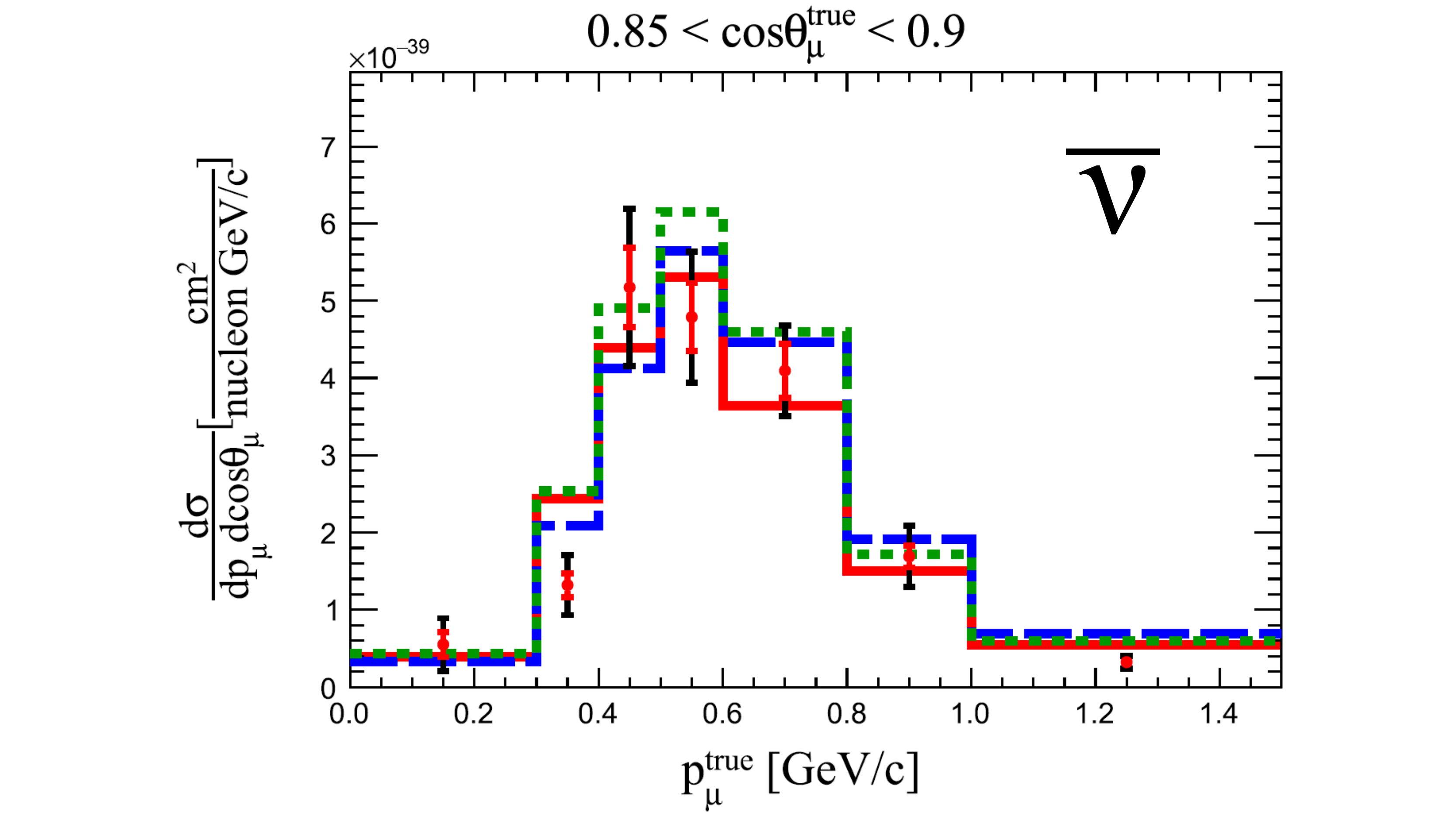}  
  \caption{Example of T2K flux-integrated $\textrm{CC0}\pi$ neutrino and antineutrino double differential cross section for a given muon scattering angle bin as a function of the muon momentum. The experimental T2K data are compared to the NEUT predictions using the RPA+2p-2h model of the Valencia group \cite{Nieves2011} (solid red line), to the RPA+np-nh calculations of the Lyon group \cite{MECM2009} (dashed blue line) and to the SuSAv2+MEC model \cite{Megias:2016fjk} (green dashed line). The figure is extracted from Ref. \cite{Abe:2020jbf}
  }
  \label{fig_cross_sections_nu_anti}
\end{figure}

In principle, if the calculations have been done consistently, all approaches should give the same cross sections. However although the trends are similar, some significant differences remain, including even differences between calculations of the same category, such as the ones of the Lyon and Valencia groups. These differences are illustrated for instance in Ref.\cite{Katori:2016yel}, where a comparison is shown between data and predictions of models which calculate several neutrino and antineutrino MiniBooNE, T2K and MINERvA flux-integrated differential cross sections. Moreover the approximations made by the different groups in the treatment of 2p-2h excitations are also discussed in this work. A more recent comparison between the different predictions and the data is published in Ref.\cite{Abe:2020jbf}. In this work the T2K flux-integrated double differential cross sections with one muon (or antimuon), zero pion and any number of nucleons, called $\textrm{CC0}\pi$ cross sections, for muon neutrino and antineutrino are given, as well as their combinations. 
An example of the published results is shown in Fig. \ref{fig_cross_sections_nu_anti}. 
From Refs.\cite{Katori:2016yel,Abe:2020jbf} it turns out that the relative 2p-2h contribution for neutrino and antineutrino cross sections is different in the different approaches and as a consequence also in their combinations. To understand these discrepancies is important for the interpretation of experiments aimed at the determination of the CP violating phase, such as the presently running T2K \cite{Abe:2019vii} and NOvA \cite{Acero:2019ksn} and the future Hyper-K \cite{Abe:2015zbg} and DUNE \cite{Acciarri:2016crz}. 

The implementation of the different theoretical approaches in the main Monte Carlo neutrino event generators, such as GENIE \cite{Andreopoulos:2009rq}, NEUT \cite{Hayato:2009zz} and NuWro \cite{Golan:2012wx} is ongoing. In this connection a delicate point is that in some models, such as the independent particle models, nucleon-nucleon SRC contributions are considered as part of two-body currents, while in other approaches, such as those starting from correlated nuclear wave functions, these SRC contributions are considered belonging to the one-body current. There is a risk of a double counting in the Monte Carlo if different contributions to the neutrino cross sections are taken from different models. 

In this paper we focus on the treatment of nuclear correlations leading to 2p-2h excitations. 
In particular we illustrate, via formal many-body techniques, the correspondence between independent particle models and correlated wave functions approaches.  
Correspondences between the different approaches up to the quasielastic region have already been analyzed in Refs. \cite{Nieves:2017lij,Sobczyk:2017mts}, largely based on the formalism of Ref.\cite{Nieves2011}, in which muon capture, pion capture, electron and neutrino cross sections were discussed. The way of connecting the scaling with short-range correlation has been discussed in Ref.\cite{Berardo:2011fx}. 
In the present article, which is somewhat of pedagogical nature, we establish the explicit connection between the dressed nucleon moving in a mean field and the bare correlated nucleon, for which the coupling to the external probe is known. 
The technique is based on unitary transformation mapping the Fermion operators relative to the bare nucleon into quasi-particle operators relative to dressed nucleon. With this technique we derive explicit formulas for spectral functions, response functions, for the momentum distribution, the separation energy, general enough to be applied with any kind of effective nucleon-nucleon interaction. 
With these tools we can evaluate whether approaches based on effective interactions, such as the ones used for the calculations of neutrino-nucleus scattering \cite{MECM2009,Nieves2011}, are compatible with the expected amount of correlations coming from \textit{ab initio} microscopic calculations (i.e. based on correlated wave functions). Our method to check this compatibility is to evaluate the Fermi sea depopulation, the separation energy and the kinetic energy per nucleon, quantities which have been debated in the context of the interpretation of the EMC effect. 

%\GC{The present article  is somewhat of a pedagogical nature, the many body effects of the various response functions which enter the neutrinos cross section are evaluated with the constraints of certain physical quantities such as the depopulation of the Fermi sea. Moreover for the delta which also enters  the many body effects of the response it is introduced on the same footing as the nuclear excitations: in this way  certain interference terms naturally appear in our formalism. }
%%%%%%%%%%%%%%%%%%%%%%%%%%%%%%%%%%%%%%%%%%%%%%%%%%%%%%%%%%%%%%%%%%%%%%%%%%%%%�
\section{Neutrino-nucleus cross-section, response functions and spectral function: generalities  }\label{Response}
%%%%%%%%%%%%%%%%%%%%%%%%%%%%%%%%%%%%%%%%%%%%%%%%%%%%%%%%%%%%%%%%%%%%%%%%%%%%%%%%%
The neutrino-nucleus double differential cross section  for the charged current reaction
\mbox{$ \nu_l \, (\bar{\nu}_l) + A \longrightarrow l^- \, (l^+) + X $}
is given by
\begin{equation}
\label{m_eq_1}
\frac{d^2 \sigma}{d\Omega_{k'} d \omega} =  
\frac{G_F^2 \cos^2\theta_C}{32\pi^2}\frac{|{\bf{k}}'|}{|{\bf{k}}|}L_{\mu\nu}W^{\mu\nu}(\omega,{\bf{q}}). 
\end{equation}
Here $d\Omega_{k'}$ is the differential solid angle in the direction specified
by the charged lepton momentum ${\bf{k}}'$ in the laboratory frame,
$\omega=E_\nu-E_l'$ is the energy transferred to the nucleus,
the zero component of the four momentum transfer $Q=K-K^\prime\equiv(\omega,{\bf{q}})$,
with $ K \equiv (E_\nu,{\bf{k}})$ and $ K^\prime \equiv (E_l',{\bf{k}}')$, being the initial and final lepton four momenta.
In Eq.~(\ref{m_eq_1}), $ G_F $ is the weak coupling constant,
$ \theta_c $ is the Cabbibo angle, and $L$ and $W$ are the leptonic and hadronic tensors, respectively.

The hadronic tensor, which describes the hadronic part, is defined as
\begin{equation}
\label{eq_Wmunu}
W^{\mu\nu}(\omega,{\bf{q}})=\sum_f
\langle \Psi_i |J^{\mu}(Q)|\Psi_f\rangle
\langle \Psi_f |J^{\nu}(Q)|\Psi_i\rangle
\delta(\omega-E_f+E_i) \,,
\end{equation}
where $|\Psi_i\rangle$ and $|\Psi_f\rangle$ are the initial and final hadronic states with energy $E_i$ and $E_f$. The electroweak nuclear current operator $J^{\mu}$ can be expressed as a sum of
one-body $J^{\mu}_{OB}$ and two-body $J^{\mu}_{TB}$ contributions. 
We remind that there are several contributions to the two-body current $J^{\mu}_{TB}$: in the electromagnetic case, these are the so called pion-in-flight term $J^{\mu}_{\pi}$,
the contact term $J^{\mu}_{\textrm{contact}}$ and the $\Delta$-intermediate state or $\Delta$-MEC term $J^{\mu}_{\Delta}$. 
At level of terminology,
in the past some authors refer just to the first two terms as Meson Exchange Currents contributions
(such as in~\cite{MECM2009}) but actually the most current convention consists of including the $\Delta$-term into MEC. In the electroweak case another contribution, the pion-pole term $J^{\mu}_{\textrm{pole}}$, appears. 
It has only the axial component and therefore it is absent in the electromagnetic case. 
%In the 2p-2h sector, the three microscopic models of Refs. \cite{MECM2009,Nieves2011,Megias2015} are based on the Fermi gas, which is the simplest independent particle model. In other words, the calculations are performed in a basis of uncorrelated nucleons. 
In the framework of independent particle models, such as Fermi gas based models or mean field based models, the calculations are performed in a basis of uncorrelated nucleons. When such a basis is used in the 1p-1h sector, 
one needs to add the nucleon-nucleon short-range correlations (NN-SRC) contributions. The way to include these contributions is through the introduction of an additional two-body current,
the correlation current $J^{\mu}_{\textrm{NN-SRC}}$.
This additional term is absent in the approaches which start from the description of the nucleus in terms of correlated wave functions 
(such as the approaches based on ``exact'' spectral function)
since the matrix elements of the one body current $\langle \Psi_i |J^{\mu}_{OB}(Q)|\Psi_f\rangle$ already includes this contribution. 
%In the approaches of the second type based on an ``exact'' spectral function, the various response functions are often calculated using a factorization approximation.
The aim of the present article is to discuss the connections between the two approaches in the treatment of the correlations leading to 2p-2h excitations. We will also show which are the contributions of $\Delta$-MEC and of the interference between the correlation and MEC which can be naturally included in the formalism. 

For this purpose, let us start by considering the response to a probe (such as the vector boson related to charged current neutrino interaction) which transfers to the nucleus an energy momentum $(\omega,{\bf q})$ and couples to the individual nucleons with operators 
$
{\cal{O}}(j)= \tau_j^\pm, \,\,\, ( {\bf{\sigma}}_j \cdot \widehat{q} ) \, 
\tau_j^\pm, \,\,\, 
( {\bf{\sigma}}_j \times \widehat{q} )^i
\, \tau_j^\pm$\footnote{The $ \sigma\tau$ operators are replaced by the usual 1/2 to 3/2 transition operators $ST$ in the case of coupling to the $\Delta$.}. This response is defined as: 
\begin{equation}
\label{eq_def_resp}
R(\omega,{\bf q})=\sum_{n}\,|< n|\sum_{j=1}^A\,{\cal{O}}(j)\, e^{i{\bf q}\cdot{\bf x}_j} |0 >|^2\,\delta(\omega-E_n + E_0), 
\end{equation}
where $|n>$ and $E_n$ are the eigenstates and the eigenvalues of the full nuclear Hamiltonian. 
The connection between the definition of the nuclear response and the definition of hadronic tensor of Eq. (\ref{eq_Wmunu}) is evident.
Very generally, the response function is the imaginary part of a two-body Green's function, the polarization propagator. In the factorization approximation 
%consists in disconnecting the particle line and the hole line, i.e, 
the two-body Green's function is factorized as a product of two one-body Green's functions, i.e, two single nucleon propagators. 
The derivation is given in the appendix \ref{appendix_factorization} with the  result, which is intuitive:
\begin{eqnarray}
R(\omega,{\bf q})&=&\sum_{k'_1, k'_2,k_1,k_2} <k'_1|{\cal{O}}^\dagger  e^{-i{\bf q}\cdot{\bf x}}|k'_2><k_2|{\cal{O}} e^{i{\bf q}\cdot{\bf x}}|k_1>\nonumber\\
&&\int_{-\infty}^{\varepsilon_F} dE_1 \, S^h_{k_1,k'_1}(E_1)\,\int_{\varepsilon_F}^{\infty} dE_2\,S^p_{k'_2,k_2}(E_2)\,\delta(\omega-E_2+E_1).
\label{Facresp}
\end{eqnarray}
The indices $k$ ($k=({\bf k},s,t)$) in nuclear matter) stand for a basis of single particle states and  $a_k$, $a^\dagger_k$, are the associated destruction and creation operators with discrete normalization $\left\{a_k,a^\dagger_{k'}\right\}=\delta_{kk'}$. This factorized response function is completely determined by the hole and particle spectral functions:
\begin{eqnarray}
S^h_{k,k'}(E)&=&\sum_n <0|a^\dagger_k|n>< n|a_{k'}|0>\,
\delta\left(E+E_n^{A-1}-E_0^A\right),  \nonumber\\ 
S^p_{k',k}(E)&=&\sum_n <0|a_{k'}|n>< n|a^\dagger_k|0>\,
\delta\left(E-E_n^{A+1}+E_0^A\right), \label{Spectral} 
\end{eqnarray}
$A$ representing the number of nucleons of the system.

%%%%%%%%%%%%%%%.%%%%%%%%%%%%%%%%%%%%%%%%%%%%%%%%%%%%%%%%%%%%%%%%%%%%%%%%%%%%%%�
  \section{Back to the EMC effect \label{sec_emc}  }
%%%%%%%%%%%%%%%%%%%%%%%%%%%%%%%%%%%%%%%%%%%%%%%%%%%%%%%%%%%%%%%%%%%%%%%%%%%%%%%%
 Before presenting a more detailed treatment of the main quantities discussed in this paper, %which  focus on nuclear correlations, 
 we would like to remind that similar problem on the role and of the treatment of correlations, discussed here in connection with neutrino scattering, already happened  thirty years ago on the interpretation of the EMC effect \cite{EMC84,EMC88}.  It concerns the structure function $F_2(x)$ which measures the probability for a quark inside the nucleon to carry a fraction
$x$ of the nucleon momentum. The EMC effect is the observed depletion of the nuclear structure function $F_{2A}(x)/A$ with respect to the free nucleon one $F_{2N}(x)$, the effect being maximal for a value  $x\simeq 0.6$ of the Bjorken variable.  In this problem  Fermi motion and binding corrections should be applied: the quark distribution in nuclei is obtained  as a convolution of the quark distribution of the bound nucleon with the momentum distribution (i.e. Fermi motion) of the nucleon in the nucleus. A simple approximate formula can be established in the relevant $0.3<x<0.8$ domain: 
\begin{equation}
\frac{1}{A}\frac{F_{2A}(x)}{F_{2N}(x)}=1\,+\,\frac{1}{M_N}\left\langle \,\epsilon\,\right\rangle\left(\frac{-xF'_{2N}(x)}{F_{2N}(x)}\right).
\label{de XXX}
\end{equation}
where $\left\langle \epsilon\right\rangle$ is the separation energy, a negative quantity which is  the opposite of the mean energy needed to remove a nucleon from the nucleus. Its expression: 
\begin{equation}
\left\langle\, \epsilon\,\right\rangle=\frac{1}{A}\int dE\int\frac{d{\bf k}}{(2\pi)^3} E \,S^h(E,{\bf k})
\label{separation}
\end{equation}
 involves the hole spectral function 
$S^h(E)$, the diagonal part of Eq.(\ref{Spectral}), which is assumed to be independent of the spin-isospin state:
\begin{eqnarray}
S^h(E,{\bf k})&=&\sum_n \left|\left\langle n\right|a_{\bf k}\left|0\right\rangle\right|^2\,
\delta\left(E+E_n^{A-1}-E_0^A\right)  \nonumber\\  
   &=& \sum_n \left|\left\langle n\right|a_{\bf k}\left|0\right\rangle\right|^2\,
\delta\left(E-\left(\left(E_0^A-E_0^{A-1}\right)- \left(E_n^{A-1}-E_0^{A-1}\right)\right)\right). 
\end{eqnarray}
The formula  above, as most  formulas of this article, is written for a piece of  nuclear matter with a given density $\rho$. It can be adapted to the nuclei within a semi-classical Thomas-Fermi approximation as in Ref. \cite{Schuck89}.
The hole spectral function is non vanishing for energy below the Fermi energy $\epsilon_F=E_0^A-E_0^{A-1}$ and its integral gives the occupation number of the state labelled by $k=({\bf k},s,t)$:
\begin{equation}
n_{\bf k}=\int_{-\infty}^{\epsilon_F} dE\, S^h(E,{\bf k}).
\end{equation}
 For typical nuclei from  $^{12}$C to $^{56}$Fe, the value $\left\langle\, \epsilon\,\right\rangle=-(40 \div 50)\,MeV$  gives a correct fit to the EMC effect. The first binding model calculation to explain the observed depletion has been performed in a shell model framework \cite{Aku85}. In such a picture, or in any Hartree-Fock (HF) scheme without density dependent forces, one has :
\begin{equation}
\frac{E_0^A}{A} \equiv -B =\frac{1}{2}\left(\left\langle\, \epsilon\,\right\rangle\,+ \left\langle\, t\,\right\rangle \right).
\label{shellmodel}
\end{equation}
Taking for the binding energy per nucleon $B=8\,MeV$ and for the kinetic energy per nucleon a value   $\left\langle\, t\,\right\rangle=20 \div 23\, MeV$, one obtains, $-\left\langle\, \epsilon\,\right\rangle=36 \div 39\,MeV$ which seems to explain most of the EMC effect. However as pointed out by G.F. Li, K.F. Liu and G.E. Brown \cite{LI88}, this conclusion is misleading since  such a description cannot reproduce both the binding energy and the experimental single particle energies. The solution of this contradiction %problem
necessitates the introduction of density dependent interactions or three-body forces. In such an approach, the energy per nucleon incorporates two-body and three-body contributions:
\begin{equation}
\frac{E_0^A}{A} \equiv -B=\left\langle\, t\,\right\rangle\,  +\,\frac{1}{2} V_2(\rho)\, + 2\,\delta V_3(\rho),
\end{equation}
where the factor two in front of the three-body term is introduced for convenience. The separation energy, which is the mean value of the single particle energy, takes the form:
\begin{equation}
\left\langle\, \epsilon\,\right\rangle =\left\langle\, t\,\right\rangle \, +\, V^{HF}\, \equiv \left\langle\, t\,\right\rangle\,  +\, V_2(\rho)\, +\, 6\, \delta V_3(\rho).
\end{equation}
 Consequently Eq. (\ref{shellmodel}) is modified according to \cite{Meyer91,Chanfray91} :
\begin{equation}
\frac{E_0^A}{A} \equiv -B =\frac{1}{2}\left(\left\langle\, \epsilon\,\right\rangle\,+ \,\left\langle\, t\,\right\rangle \right) \,-\,\delta V_3.
\label{skyrme}
\end{equation}
It contains a contribution coming from the effective three-body term, the so-called rearrangement term, $\delta V_3$.
Using a  $SIII$ Skyrme interaction one obtains for $^{56}$Fe,  $B=8.5\, MeV$, $ \left\langle\, t\,\right\rangle =18\, MeV$, $\delta V_3=6\,MeV$ \cite{Meyer91}. It follows that $\left\langle\, \epsilon\,\right\rangle = -23\, MeV$, which is much too small to reproduce the EMC effect. From these considerations, the authors of Ref. \cite{LI88} concluded that conventional nuclear models are unable to reproduce the EMC effect, the major part coming from other sources possibly related to an intrinsic modification of the nucleon structure in the nuclear medium. This conclusion as well was  premature, as shown below. 

To see this point, let us now start from a bare nucleon-nucleon interaction, at variance with the previous one like Skyrme  or Gogny forces which are effective ones corresponding more to a $G$ matrix. As observed by M. Ericson \cite{Ericson87} who first pointed out the role of the correlations in the EMC effect, the separation energy $\left\langle\, \epsilon\,\right\rangle$ can be obtained from Eq. (\ref{shellmodel}) which is not limited to the simplest Hartree-Fock scheme but constitutes an exact result known as the Koltun sum rule:    
\begin{equation}
-\left\langle\, \epsilon\,\right\rangle=\left\langle\, t\,\right\rangle +2\,B.
\label{koltun}
\end{equation}
Here the separation energy is the one defined from the spectral function of Eq. (\ref{separation}). The important point is therefore that $\left\langle\, t\,\right\rangle$ is the real kinetic energy of the bare nucleons contrary to the Hartree-Fock quantity appearing in Eqs. (\ref{shellmodel}), (\ref{skyrme}) where, as will be precised in section \ref{Landau}, the "effective HF nucleons" are dressed objects. Realistic calculations based on correlated wave functions have shown that the bare nucleon momentum distribution acquires a long tail beyond Fermi momentum since the effect of short-range correlation is to depopulate the Fermi sea. For instance, in the evaluation of Ciofi degli Atti et al. \cite{Cioffi}, for nuclei ranging from $^{12}$C to $^{56}$Fe, the kinetic energy per nucleon becomes larger, namely $\left\langle\, t\,\right\rangle\simeq 35\,MeV$ and the absolute value of the separation energy as well becomes  larger,  $-\left\langle\, \epsilon\,\right\rangle\simeq 50\,MeV$. Thus conventional nuclear effects are indeed able to reproduce the major part of the EMC effect.  One can  conclude that the independent particle picture is a very approximate view of the nucleus or that the Hartree Fock scheme  with bare nucleons is a poor approximation of the nuclear ground state. This conclusion looks surprising at first since we know that independent particle models based on the HF scheme reproduce accurately many basic properties of nuclei. The reason of the apparent discrepancy is due  to the implicit identification of the two degrees of freedom: the bare and the dressed nucleons. In reality the Hartree-Fock nucleons should not be treated as bare nucleons but as nucleons surrounded by a polarization cloud, constituting  quasi-particles (in the Landau sense), moving independently in a smooth mean field potential. As pointed out by \cite{Chanfray91} such dressed objects couple differently than the bare ones to the external (electroweak) probe. The problem of the explicit construction of these dressed objects from the bare ones has rarely been adressed in the literature with some exceptions \cite{Desplanques,Chanfray91}. We will come with more details to this question, a central aspect of the present article, in the next section.
%%%%%%%%%%%%%%%.%%%%%%%%%%%%%%%%%%%%%%%%%%%%%%%%%%%%%%%%%%%%%%%%%%%%%%%%%%%%%%�
\section{Dressed nucleons versus bare nucleons: Landau quasi-particles}
\label{Landau}
%%%%%%%%%%%%%%%%%%%%%%%%%%%%%%%%%%%%%%%%%%%%%%%%%%%%%%%%%%%%%%%%%%%%%%%%%%%%%%%%%
\subsection{Explicit construction of the dressed nucleons from the bare nucleons}
The previous discussion shows that it is highly desirable to establish a connection between the descriptions of the nucleus either with independent HF dressed nucleons or with correlated bare nucleons for which we know the coupling to the external probe. 
%This problem of the explicit construction of the dressed nucleon states is rarely addressed in the literature. 
Here we follow a method proposed by B. Desplanques \cite{Desplanques} for the case of nuclear matter. 

Let us consider a system described by an Hamiltonian where the bare nucleon-nucleon interaction reduces to a two-body potential. According to Goldstone \cite{Goldstone}, the correlated ground state of the nucleus, $|0>$, can be obtained from the state of uncorrelated bare nucleons by a unitary transformation:
\begin{eqnarray}
|0>&=&U\, |0>_{uncorr} = U(a) \,\left(\prod_h a_h^\dagger\right)\,|vac>=U(a) \,\left(\prod_h a_h^\dagger\right)\,U^\dagger(a)\,|vac>\nonumber\\
&\equiv& \left(\prod_h A_h^\dagger\right)\,|vac>,
\end{eqnarray}
where $|vac>$ is the true vacuum state with baryonic number equal to zero. The labels $h$ stand for hole states with momentum below $k_F$.
Hence the correlated ground state $|0>$ can be obtained as a Slater determinant of dressed nucleons created by the $A_h^\dagger$ operators. These operators are 
related to the bare creation operators, $a_h^\dagger$, by a unitary transformation preserving the anticommutation relation: 
\begin{equation}
A_h^\dagger=U(a)\, a_h^\dagger\, U^\dagger(a)\,\,\,\Leftrightarrow\,\,\,a_h^\dagger=U^\dagger(A)\, A_h^\dagger\, U(A).
\end{equation}
The operator $U(A)$ is  obtained by requiring that the Hamiltonian, $$H(a)\equiv U^\dagger(A)\, H(A) \,U(A),$$ admits the correlated ground state $|0>$ as an eigenstate. Hence, all terms such as $A_{p_1}^\dagger A_{p_2}^\dagger A_{h_2} A_{h_1}$  should disappear from the Hamiltonian when it is expressed in term of the dressed operators $A, A^\dagger$.
Obviously this construction can be done only within some approximations. 
%NOUVEAU POUR REPONDRE AU REFEREE
Starting, as in Ref.\cite{Desplanques}, with an unitary operator $U(A)=\exp(S(A))$, $S(A)$ being an anti-hermitic operator truncated at the 2p-2h excitations 
\begin{equation}
S(A)=\frac{1}{2}\left(\alpha_{p_1 p_2 h_2 h_1} A^\dagger_{p_1}A^\dagger_{p_2}A_{h_2}A_{h_1}\,-\,hc\right),
\end{equation}
one obtains the following result  for hole and particle creation and annihilation operators :
\begin{eqnarray}
%\label{eq_ah}
a_h &=& \sqrt{Z_h}\,A_h \,-\,\frac{1}{2}\,\sum_{h_2 p_3 p_4}\frac{<p_3\, p_4\,|\,\bar{G}\,|\,h_2\, h>}{\epsilon_h +\epsilon_{h_2}-\epsilon_{p_3}-\epsilon_{p_4 }}\,A^\dagger_{h_2}\,A_{p_4} \,A_{p_3} \label{eq_ah}\\
a_p &=& \sqrt{Z_p}\,A_p \,-\,\frac{1}{2}\,\sum_{p_2 h_3 h_4}\frac{<p\, p_2\,|\,\bar{G}\,|\,h_3\, h_4>}{\epsilon_p +\epsilon_{p_2}-\epsilon_{h_3}-\epsilon_{h_4 }}\,A^\dagger_{p_2}\,A_{h_3} \,A_{h_4},
\label{eq_ap}
\end{eqnarray}
where $Z_h$ and  $Z_p$ are normalization factors which are fixed to ensure that $<0|\{a_k,a^\dagger_k\}|0>=1$ :
\begin{eqnarray}
Z_h &=& 1\,-\,\frac{1}{2}\,\sum_{h_2 p_3 p_4}\left|\frac{<p_3\, p_4\,|\,\bar{G}\,|\,h_2\, h>}{\epsilon_h +\epsilon_{h_2}-\epsilon_{p_3}-\epsilon_{p_4 }}\right|^2\label{Zh}
\label{hZfactor}
\\
Z_p &=& 1 \,-\,\frac{1}{2}\,\sum_{p_2 h_3 h_4}\left|\frac{<p\, p_2\,|\,\bar{G}\,|\,h_3\, h_4>}{\epsilon_p +\epsilon_{p_2}-\epsilon_{h_3}-\epsilon_{h_4 }}\right|^2 \label{Zp}.
\label{Zfactor}
\end{eqnarray}
In the above expressions (Eqs. \ref{eq_ah},\ref{eq_ap},\ref{Zh},\ref{Zp}) the energies appearing in energy denominators are the Hartree-Fock energies and $\bar G$ has to be identified with the (antisymmetrized) on-shell G-matrix. It can be checked that the part of the Hamiltonian acting on the ground state is:
\begin{equation}
H_{gs}=\sum_{hh'}<h\,|\,T\,|\, h'> A^\dagger_{h}A_{h'} +\frac{1}{4} \sum_{h_1 h_2 h'_1 h'_2}<h_1\,h_2\,|\,\bar{G}\,|\,h'_1\, h'_2>
A^\dagger_{h_1}A^\dagger_{h_2}A_{h'_2}A_{h'_1}.
\end{equation}
Consequently the ground state energy is the one expected from using the standard on-shell G matrix, namely:
\begin{equation}
E_0= \sum_{h}<h\,|\,T\,|\, h>  +\frac{1}{2}  \sum_{h_1 h_2}<h_1\, h_2\,|\,\bar{G}\,|\,h_1\, h_2>.
\end{equation} 
The important point is that this $G$  matrix, often identified with the effective interaction in Hartree-Fock calculation, refers to objects, the dressed or renormalized nucleons, which are definitely different from the bare nucleons. \\
If an energy independent mean field is assumed, then the single particle energy difference, entering for example in Eqs.(\ref{eq_ah}) and (\ref{eq_ap}), involves only the kinetic energy difference between particle and hole states. 
%%%%%%%%%%%%%%%%%%%%%%%%%%%%%%%%%%%%%%%%%%%%%%%%%%%%
\subsection{The spectral functions}
%%%%%%%%%%%%%%%%%%%%%%%%%%%%%%%%%%%%%%%%%%%%%%%%%%%%%%%%%
The states $|n>$ appearing in the expression of the hole (particle) spectral function of Eq. (\ref{Spectral}) are either a one-hole (one particle) state or a 2 hole-1 particle (2 particle-1 hole) state obtained from the ground state $|0>$ by action of dressed creation ($A^\dagger$) and annihilation ($A$) operator.
Notice that the $\Delta$-resonance states can be viewed as particular particle states above the Fermi energy, hence they can be treated on the same footing as nucleon states in the calculations of the response function. Using the notation $\rho_k=\Theta\left(\epsilon_F-\epsilon_k\right),\, \bar\rho_k= \Theta\left(\epsilon_k - \epsilon_F\right)$, the explicit form of these spectral functions can be straightforwardly obtained as:
\begin{eqnarray}
S^h_{k,k'}(E)&=&\sum_n <0|a^\dagger_k|n>< n|a_{k'}|0>\,
\delta\left(E+E_n^{A-1}-E_0^A\right) = Z_k \,\delta_{kk'}\, \delta\left(E-\epsilon_k\right)\,\rho_k\nonumber\\ 
&+&\frac{1}{2}\,\sum_{p_2 h_3 h_4}\frac{<k\, p_2\,|\,\bar{G}\,|\,h_3\, h_4>\,
<k'\, p_2\,|\,\bar{G}\,|\,h_3\, h_4>}{\left(\epsilon_k +\epsilon_{p_2}-\epsilon_{h_3}-\epsilon_{h_4 }\right)\,
\left(\epsilon_{k'} +\epsilon_{p_2}-\epsilon_{h_3}-\epsilon_{h_4}\right)}\nonumber\\ &&
\delta\left(E-\left[\epsilon_{h_3}+\epsilon_{h_4}-\epsilon_{p_2}\right]\right)\,\bar\rho_k\,\bar\rho_{k'}
\label{Spectral3}
\end{eqnarray}
\begin{eqnarray}
S^p_{k',k}(E)&=&\sum_n <0|a_{k'}|n>< n|a^\dagger_k|0>\,
\delta\left(E-E_n^{A+1}+E_0^A\right)\,= Z_k \,\delta_{kk'} \,\delta\left(E-\epsilon_k\right)\,\bar\rho_{k}\nonumber\\ 
&+&\frac{1}{2}\,\sum_{h_2 p_3 p_4}\frac{<p_3\, p_4\,|\,\bar{G}\,|\,h_2\,k'>
<p_3\, p_4\,|\,\bar{G}\,|\,h_2\,k>}{\left(\epsilon_k +\epsilon_{h_2}-\epsilon_{p_3}-\epsilon_{p_4 }\right)\,
\left(\epsilon_{k'} +\epsilon_{h_2}-\epsilon_{p_3}-\epsilon_{p_4}\right)}\nonumber\\
&&\delta\left(E-\left[\epsilon_{p_3}+\epsilon_{p_4}-\epsilon_{h_2}\right]\right)\,
\rho_k\,\rho_{k'}.
 \label{Spectral2} 
\end{eqnarray}
In nuclear matter these spectral functions have to be diagonal in momentum space, i.e., ${\bf k}={\bf k}'$. Consequently if we consider only nucleon degrees of freedom, these spectral functions are strictly diagonal. At the level we are, this property  remains true in presence of the $\Delta$ for the particle spectral function since $\epsilon^\Delta_k\simeq \omega_\Delta + (k^2/2 M_\Delta)+U^{MF}_\Delta$ 
(with $\omega_\Delta=M_\Delta-M_N$ and the mean field $U^{MF}$ has to be identified with the Hartree-Fock single particle potential)
remains always larger than the Fermi energy, $\epsilon_F=(k^2_F/2 M_N)+U_N^{MF}$. 

However one can have non diagonal, i.e. $N\Delta$,  interference contributions in the hole spectral function with $k$ being a nucleon state and $k'$ being a $\Delta$ state with the same momentum. This is an important feature  since it opens the possibility of generating the $N\Delta$ interference piece of the spin-isospin responses, as in Refs. \cite{MECM2009}, \cite{AEM84}.

Beyond the spectral function, another important quantity is the mass operator, which is crucial for the treatment of exclusive processes. 
Using Eq.(\ref{eq_appendix_im_M}) of appendix \ref{appendix_G_and_M} one obtains for its imaginary part:  
\begin{eqnarray}
-\frac{1}{\pi} Im \,M(E,{\bf k})&=&-\frac{1}{2\pi}\bigg[\sum_{h_2 p_3 p_4}\left|<p_3\, p_4\,|\,\bar{G}\,|\,h_2\,k>\right|^2\,\rho_{h_2}\,\bar\rho_{p_3}\,\bar\rho_{p_4}\nonumber\\
&&\qquad\delta\left(E-\left[\epsilon_{p_3}+\epsilon_{p_4}-\epsilon_{h_2}\right]\right)\nonumber\\
&&-\,\sum_{p_2 h_3 h_4}\left|<k\, p_2\,|\,\bar{G}\,|\,h_3\, h_4>\right|^2\,\bar\rho_{p_2}\,\rho_{h_3}\,\rho_{h_4}\nonumber\\
&&\qquad\delta\left(E-\left[\epsilon_{h_3}+\epsilon_{h_4}-\epsilon_{p_2}\right]\right)\bigg]\label{Mass},
\end{eqnarray}
which is the usual form given in particular in Ref. \cite{Schuck89}, apart from the fact that in this paper the $G$ matrix is not the on-shell one but the in-medium $T$-matrix with  its full energy dependence. However, as discussed in \cite{Schuck89}, for many nuclear physics application it is not necessary to start from scratch, i.e. from the bare nucleon-nucleon interaction. It is sufficient to 
identify the $G$ matrix with some effective forces adjusted on nuclear ground states properties. One can thus replace  $G$, as in \cite{Schuck89}, by the Gogny force or by the Skyrme force. One can also replace $\bar G$, by the effective interaction containing pion and rho exchange in presence of short range correlations described through the introduction of Landau-Migdal $g'$ parameters. In this way one makes an explicit connection, one motivation of the present work, with the calculations of Refs. \cite{MECM2009,Nieves2011} on neutrino scattering. One can also replace, as we will do in section \ref{subsec_explicit_calc_Dn_T}, $\bar G$ by effective interactions built on our chiral effective theory \cite{CE2007,CE2011,MC2008,MC2009} or similar approaches developed by the Osaka group \cite{Hu2009,Hu2010,Hu2011}.\\
%%% NOUVEAU PARAGRAPH POUR REPONDRE AU REFEREE
The above spectral functions will be used in the following to estimate the amount of correlations through the Fermi sea depopulation affecting the momentum distribution. 
However, before doing that, we should stress that this approach do not possess yet the richness of the detailed calculations given in Refs.\cite{Benhar89,Benhar90,Benhar92}  in the context of the Correlated Basis Function perturbation theory. 
A one to one correspondence between the two formalisms should deserve further investigations in particular on the complete admixture of two-particle two-hole
configurations into the interacting ground state and its consequences on the one-body Green's function, essential to obtain the smooth (\textit{i.e.} non-pole) contributions of the momentum distributions \cite{Benhar20}.

%%%%%%%%%%%%%%%%%%%%%%%%%%%%%%%%%%%%%%%%%%%
\subsection{The occupation numbers}
%%%%%%%%%%%%%%%%%%%%%%%%%%%%%%%%%%%%%%%%%%%%%%%%%
\label{subsec_occupation_number}
The diagonal components ($k'=k$) of the spectral functions of Eqs.(\ref{Spectral}) once integrated on energy are related to the occupation number of the state $k$:
\begin{eqnarray}
\int_{-\infty}^{\epsilon_F} dE\, S^h_k(E) &=&<0|a^\dagger_k a_k|0>=n_k\nonumber\\
\int^{\infty}_{\epsilon_F} dE\, S^p_k(E) &=&<0|a_k a^\dagger_k |0>=1-n_k \nonumber\\
 \frac{4}{A}\sum_{\bf k} n_{\bf k}&=&\frac{3}{k^3_F}\int_0^\infty dk k^2\,n_{\bf k}=1.
\end{eqnarray}
In the last line, we have explicitly shown the normalization by factorizing the spin-isospin factor (4 for nuclear matter) and by introducing the quantity $n_{\bf k}\equiv n_k$ which depends only on three-momentum ${\bf k}$. 
In general, the occupation number can be decomposed according to :
\begin{equation}
\label{eq_nk_decomposition}
n_{\bf k}=\Theta (k_F-k) \left(1-\Delta n^h_{\bf k}\right) +\Theta (k-k_F)\Delta n^p_{\bf k},
\end{equation}
where $\Delta n^h_{\bf k}$ represents the depopulation of the hole state $k$ (independently of the spin-isospin state) and $\Delta n^p_{\bf k}$ the population of the particle state $k$ originating from the correlations in the ground state.
Particle number conservation imposes: 
\begin{equation}
\frac{3}{k^3_F}\int_0^{k_F} dk k^2\,\Delta n^h_{\bf k}=\frac{3}{k^3_F}\int_{k_F}^\infty dk k^2\,\Delta n^p_{\bf k}=\Delta n, 
\end{equation}
 where $\Delta n$ represents the total depopulation (per nucleon) of the Fermi sea, which is typically $0.15\div0.2$ for light nuclei such as Carbon \cite{Cioffi}. We remind that the $\Delta$-resonance states can be viewed as particular particle states above the Fermi energy, hence  treated on the same footing as nucleon states. 

The occupation number of state $k$, which is related to the hole spectral function, writes:
\begin{eqnarray}
n_k=\int_{-\infty}^{\epsilon_F} dE\, S^h_k(E) &=&<0|a^\dagger_k a_k|0> \equiv n^h_k\,\Theta\left(\epsilon_F-\epsilon_k\right)\,+\,n^p_k\,\Theta\left(\epsilon_k - \epsilon_F\right)\nonumber\\
&\equiv& Z_{h=k}\,\Theta\left(\epsilon_F - \epsilon_k\right)\,+\,\left( 1-Z_{p=k}\right)\,\Theta\left(\epsilon_k - \epsilon_F\right)\nonumber\\
&\equiv&\left(1 - \Delta n^h_k\right)\,\Theta\left(\epsilon_F-\epsilon_k\right)\,+\,\Delta n^p_k\,\Theta\left(\epsilon_k - \epsilon_F\right)\nonumber\\
&\equiv&\left(1 - \Delta n^h_{\bf k}\right)\,\Theta\left(k_F-k\right)\,+\left[\Delta n^N_{\bf k}\,\Theta\left(k-k_F\right)
\,+\,\Delta n^\Delta_{\bf k}\right]
\end{eqnarray}
with the property
\begin{equation}
\sum_h \Delta n^h=\sum_p \Delta n^p\,\,\,\Leftrightarrow\,\,\frac{3}{k^3_F}\int_0^{k_F}  dk k^2\,\Delta n^h_{\bf k}=
\frac{3}{k^3_F}\int_{k_F}^\infty  dk k^2\,\Delta n^N_{\bf k}\,+\,\frac{3}{k^3_F}\int_0^\infty  dk k^2\,\Delta n^\Delta_{\bf k}
\end{equation}
reflecting baryon number conservation. 
In the two formulas above, $\Delta n^p_k$ has been split into a  nucleon contribution, $\Delta n^N_{\bf k}$, and a $\Delta$ one, $\Delta n^\Delta_{\bf k}$. 
Expliciting the normalization factors $Z_h$ and $Z_p$ according to Eqs. (\ref{hZfactor}),(\ref{Zfactor})  with $h=N$ and $p=N\,\hbox{or}\,\Delta$, we find for the occupation numbers: 
\begin{eqnarray}
n_h\equiv 1-\Delta n^h= &=& 1\,-\,\frac{1}{2}\,\sum_{h_2 p_3 p_4}\left|\frac{<p_3\, p_4\,|\,\bar{G}\,|\,h_2\, h>}{\epsilon_h +\epsilon_{h_2}-\epsilon_{p_3}-\epsilon_{p_4 }}\right|^2\\
n_p\equiv \Delta n^p &=& \frac{1}{2}\,\sum_{p_2 h_3 h_4}\left|\frac {<p\, p_2\,|\,\bar{G}\,|\,h_3\, h_4>} {\epsilon_p +\epsilon_{p_2}-\epsilon_{h_3}-\epsilon_{h_4 }}\right|^2,
\end{eqnarray}
which is a well known form, quoted for instance in Ref. \cite{Ramos91}, Eqs. (2.31, 2.32). 

The detailed form of these occupation numbers will be  given in subsection \ref{subsec_explicit_calc_Dn_T} for an interaction of type $\pi+\rho+g'$, the type of interaction used in the neutrino cross sections calculations of Refs. \cite{MECM2009,Nieves2011}. The specific form of the kinetic energy per nucleon, which constitutes another good indicator of the amount of correlations (see the above discussion of the EMC effect), will be  given in the same subsection, where numerical results are discussed.
%%%%%%%%%%%%%%%%%%%%%%%%%%%%%%%%%%%%%%%%%%%%%%%%%%%%%%%%%%%%%%%%%%%%%%%%
\subsection{Response function in the factorization scheme}
%%%%%%%%%%%%%%%%%%%%%%%%%%%%%%%%%%%%%%%%%%%%%%%%%%%%%%%%%%%%%%%%%%%%%%%%
The explicit form of the spectral functions, once inserted in Eq. (\ref{Facresp}),
allows to express also the response function in terms of the $Z$ factors and of the matrix elements of the $\bar G$ matrix. 
In this case, the various contributions to the response function, represented in Fig.\ref{f3}, are: 

\begin{figure}
 \centering
  \includegraphics[width=1.0\textwidth]{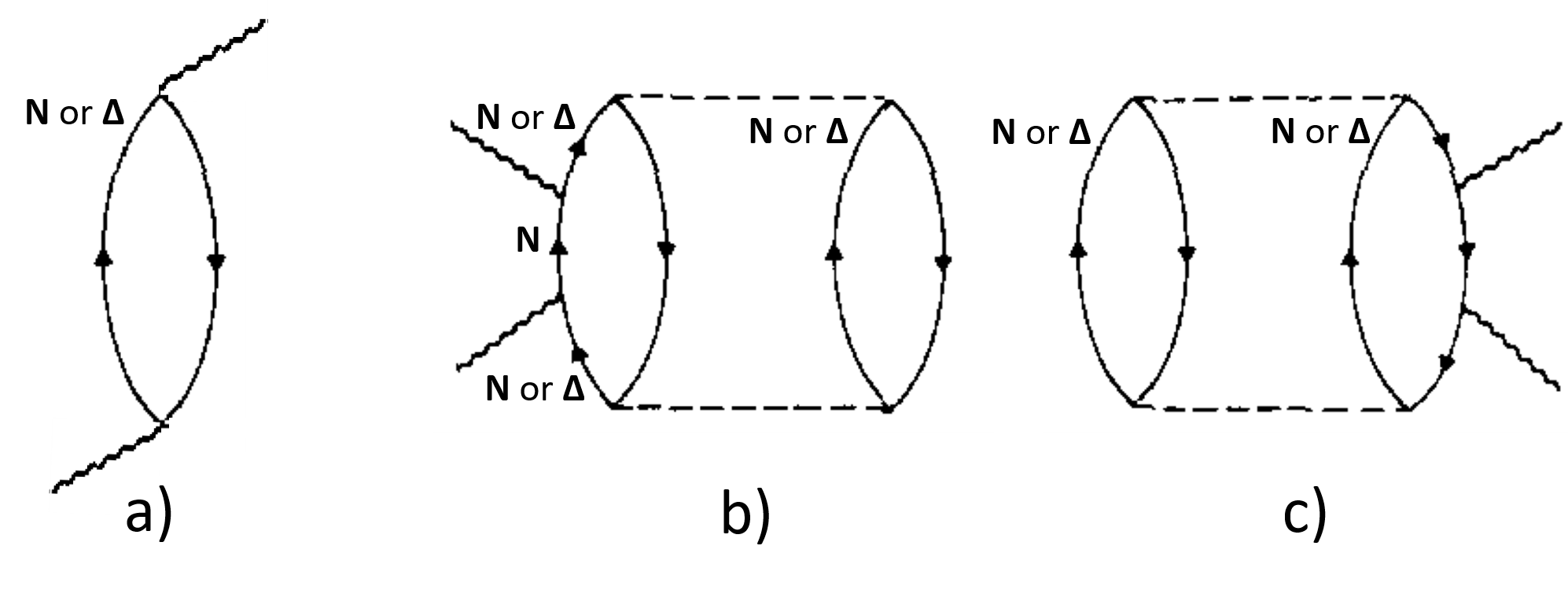}
  \caption{Feynman graphs representing various contributions of the response function: a) pure p-h contribution; b) correlation contribution; c) polarization contribution. The wiggled line represents the external probe and the dashed line the effective interaction. Continuous line with up (down) arrow represents particle (hole) state. }
  \label{f3}
\end{figure}

\begin{itemize}
\item [a)] The pure $p-h$ response quenched by the $Z$ factors. 

\begin{eqnarray}
R(\omega,{\bf q})&=&\sum_{k'_1, k'_2,k_1,k_2} <k'_1|{\cal{O}}^\dagger  e^{-i{\bf q}\cdot{\bf x}}|k'_2><k_2|{\cal{O}} e^{i{\bf q}\cdot{\bf x}}|k_1>\nonumber\\
&&\int_{-\infty}^{\varepsilon_F} dE_1 \, \,Z_{k_1}\,\rho_{k_1}\delta\left(E_1-\epsilon_{k_1}\right)\delta_{k_1,k'_1}\int_{\varepsilon_F}^{\infty} dE_2\,\,Z_{k_2}\,\bar\rho_{k_2}\delta\left(E_2-\epsilon_{k_2}\right)\delta_{k_2,k'_2}\nonumber\\
&&\delta(\omega-E_2+E_1).
\label{zerorder}
\end{eqnarray}

\item[b)] The (generalized) correlation piece obtained from the convolution of the bare particle spectral function and the dressed piece of the hole spectral function. It corresponds to the interaction of the probe with a nucleon above the Fermi sea or with a  $\Delta$ state, producing a $2h-1p$ state in the residual $A-1$ nucleus. 
This correlation piece of the response concerns not only a $NN$ state but also $N\Delta$ interference terms and $\Delta\Delta$ terms not reducible to the modification of the width of the $\Delta$ (see Fig. \ref{f3} b). The explicit expression of the correlation piece is:
%\GC{reste à écrire les expressions des différentes réponses et à faire des estimations numériques}
%\subsubsection{The correlation response including Delta} Let us now come to the correlation contribution
\begin{eqnarray}
R_\alpha^{corr}(\omega,{\bf q})&=&\sum_{k'_1, k'_2,k_1,k_2} <k'_1|{\cal{O}}_\alpha^\dagger  e^{-i{\bf q}\cdot{\bf x}}|k'_2><k_2|{\cal{O}}_\alpha e^{i{\bf q}\cdot{\bf x}}|k_1>\nonumber\\
& &\int_{\epsilon_F}^{\infty} dE_2  \,Z_{k_2}\,\bar\rho_{k_2}\delta\left(E_2-\epsilon_{k_2}\right)\delta_{k_2,k'_2}\nonumber\\
&&\frac{1}{2}\,\int_{-\infty}^{\epsilon_F} dE_1
\sum_{p_2 h_1 h_2}\frac{<k_1\, p_2\,|\,\bar{G}\,|h_1\,h_2>
<k'_1\,p_2|\,\bar{G}\,|h_1\,h_2>}{\left(\epsilon_{k_1} +\epsilon_{p_2}-\epsilon_{h_1}-\epsilon_{h_2} \right)\,
\left(\epsilon_{k'_1} +\epsilon_{p_2}-\epsilon_{h_1}-\epsilon_{h_2}\right)}\,
\nonumber\\
&&\delta\left(E_1-\left[\epsilon_{h_1}+\epsilon_{h_2}-\epsilon_{p_2}\right]\right)\,\bar\rho_{k_1}\,\bar\rho_{k_{1}'}\,\delta(\omega-E_2+E_1),
\end{eqnarray}
where we have separated the momentum integrals and the discrete spin-isospin summation. The states appearing in the matrix element of the ${\cal{O}}$  operators are spin-isospin states.
The state $k_2$ is a nucleon particle state. The states $k_1$ and $k'_1$ are either nucleon particle state or $\Delta$ state with ${\bf k}_1={\bf k}'_1$.

\item [c)] The polarization piece  obtained from the convolution of the bare hole spectral function  and the dressed piece of the particle spectral function. It corresponds to the interaction of the probe with a hole state producing a $2p-1h$ state in the residual $A-1$ nucleus.
%\subsubsection{The polarization response}
%Lets us first consider the polarization piece of the response for a coupling operaror $\cal{O}_\alpha$. Starting from the factorized form  (Eq. \ref{Facresp}) we obtain
Its explicit expression is: 
\begin{eqnarray}
R_\alpha^{pol}(\omega,{\bf q})&=&\sum_{k'_1, k'_2,k_1,k_2} <k'_1|{\cal{O}}_\alpha^\dagger  e^{-i{\bf q}\cdot{\bf x}}|k'_2><k_2|{\cal{O}}_\alpha e^{i{\bf q}\cdot{\bf x}}|k_1>\nonumber\\
& & \int_{-\infty}^{\epsilon_F} dE_1 \,Z_{k_1}\,\rho_{k_1}\delta\left(E_1-\epsilon_{k_1}\right)\delta_{k_1,k'_1}\nonumber\\
&&\frac{1}{2}\,\int_{\epsilon_F}^{\infty} dE_2
\sum_{h_2 p_1 p_2}\frac{<p_1\, p_2\,|\,\bar{G}\,|k_{2}'\,\,h_2>
<p_1\, p_2\,|\,\bar{G}\,|k_2\,h_2>}{\left(\epsilon_{k_2} +\epsilon_{h_2}-\epsilon_{p_1}-\epsilon_{p_2} \right)\,
\left(\epsilon_{k_{2}'} +\epsilon_{h_2}-\epsilon_{p_1}-\epsilon_{p_2}\right)}\, 
\nonumber\\
&&\delta\left(E_2-\left[\epsilon_{p_1}+\epsilon_{p_2}-\epsilon_{h_2}\right]\right)\rho_{k_2}\,\rho_{k_{2}'}\,\delta(\omega-E_2+E_1).
\end{eqnarray}
The state $k_1$ is a nucleon hole state. The states $k_2$ and $k'_2$ are also nucleon hole states with ${\bf k}_2={\bf k}'_2$. The states $p_1,p_2$ can be either a particle state or a $\Delta$ state.
\end{itemize}

\section{Evaluating the role of correlations}
\subsection{Sum Rule}
\label{subsection_sum_rule}
For an overview of the effect of correlations, we introduce the non-energy-weighted sum rule for the charge or the spin-isospin (ignoring the $\Delta$) response functions:
\begin{equation}
S({\bf q})=\frac{1}{A}\int d\omega\,R(\omega,{\bf q})=\frac{4}{A}\sum_{\bf k} n_{\bf k}\left(1-n_{\bf k +\bf q}\right).
\label{sumrule}
\end{equation}
For the free Fermi Gas $S({\bf q})$ reduces to:
\begin{equation}
\label{eq_S_FG}
S^{FG}({\bf q})=\Theta (q-2k_F)+\Theta (2k_F - q)\left[\frac{3}{2}\left(\frac{q}{2k_F}\right) -\frac{1}{2}\left(\frac{q}{2k_F}\right)^3\right],
\end{equation}
where $q=|\bf{q}|$ and $k_F$ is the Fermi momentum. The quantity  $S^{FG}({\bf q})$
is equal to one for $q>2k_F$ and, due to total Pauli Blocking, vanishes for $q=0$. 

According to the decomposition of Eq.(\ref{eq_nk_decomposition}), 
 the sum rule can be split in various contributions : 
\begin{eqnarray}
\label{eq_split_sum_rule}
S({\bf q}) &=&(1/A)\sum_{\bf k}\Theta (k_F-k)\,\Theta (|{\bf k +\bf q}|-k_F)\nonumber\\
&& + (1/A)\sum_{\bf k}  \Theta (k-k_F)\,\Theta (|{\bf k -\bf q}|-k_F) \,\Delta n^p_{\bf k -\bf q}\nonumber\\
&& + (1/A)\sum_{\bf k}  \Theta (k_F-k)\,\Theta (k_F-|{\bf k +\bf q}|)\, \Delta n^h_{\bf k +\bf q}\nonumber\\
&& - (1/A)\sum_{\bf k}\Theta (k_F-k)\,\Theta (|{\bf k +\bf q}|-k_F)\,\Delta n^h_{\bf k}\nonumber\\
&& - (1/A)\sum_{\bf k}\Theta (k_F-k)\,\Theta (|{\bf k +\bf q}|-k_F)\, \Delta n^p_{\bf k + \bf q}\nonumber \\
&&+ \,\hbox{second order terms in Fermi sea depletion.}
\end{eqnarray}

\begin{figure}
 \centering
  \includegraphics[width=0.8\textwidth]{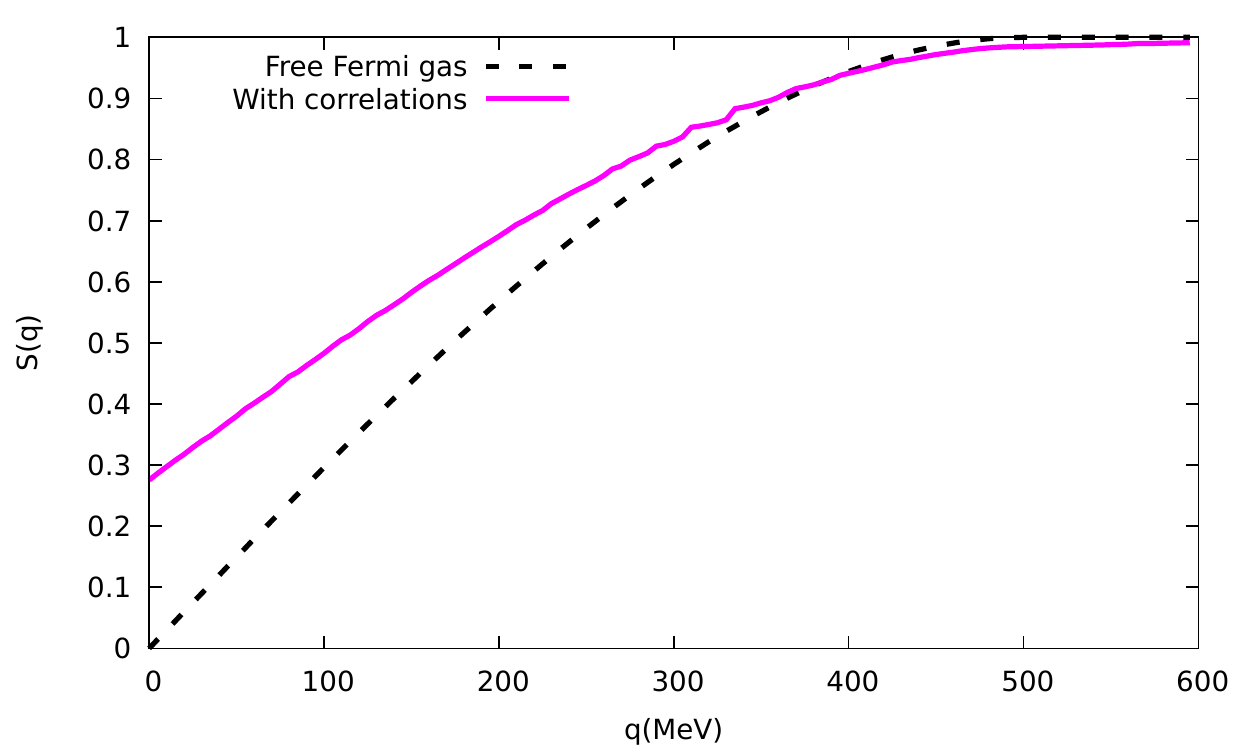}
  \caption{Sum rule as a function of momentum transfer in the free Fermi gas case and for including nucleon correlations. The value of the Fermi momentum is fixed at $k_F=245\,MeV$.}
  \label{f1}
\end{figure}

The first term is the free Fermi gas result. As for the second and third terms, 
they explicitly appear as genuine 2p-2h contributions in the expression of the responses $R^{corr}_\alpha$ and $R^{pol}_\alpha$. The second term represents the collision of the probe with a nucleon belonging to the tail of the momentum distribution (beyond the Fermi momentum), with the ejection of a particle  nucleon state; it corresponds to a correlation diagram 
of Fig. \ref {f3} b, related to $R^{corr}_\alpha$. The third term represents the collision of the probe with a nucleon in the Fermi sea but with the scattered nucleon also in the Fermi sea, which is possible since the Fermi sea is partially depopulated;  this corresponds to a  polarization diagrams 
of Fig. \ref {f3} c, related to $R^{pol}_\alpha$. 
The fourth term comes from the depletion effect ($Z_h$ factor) of the Fermi sea. 
The fifth term, instead, is associated with the $Z_p$ factor. 
%represents a depletion of the particle spectral function originating from the fact that there exists a Pauli blocking effect in the momentum tail since nucleon with momentum larger than the Fermi momentum are present in the nucleus. 
If the momentum ${\bf{q}}$ is not too large, it is the 2p-2h enhancement (second and third line of Eq.(\ref{eq_split_sum_rule})) of the sum rule which wins against the depletion effect. In particular, at zero momentum the sum rule does not vanish, 
which would be the case for the free Fermi gas, 
and, to leading order in the depletion, we find:
\begin{equation}
   S(0)\approx 2 \Delta n, 
\end{equation}
%$S(0)\approx 2 \Delta n$ 
which equally comes from the correlation (Fig. \ref{f3} b) and polarization (Fig. \ref{f3} c) diagrams. 
For the momentum evolution of this enhancement we take a simplified momentum distribution
\begin{eqnarray}
&&n_{\bf k}=\Theta (k_F-k) \left(1-\Delta n\right) +\Theta (k-k_F) \theta(\Lambda -k)\,\Delta n',\nonumber\\
&&\hbox{with the constraint:}\qquad\Delta n=\frac{3}{k^3_F}\int_{k_F}^\Lambda  dk k^2\,\Delta n',
%<T>=\frac{3}{k^3_F}\int_0^\Lambda dk k^2\,\frac{k^2}{2M}\,n_{\bf k}=35\,MeV,
\end{eqnarray}
which depends on two parameters, $\Delta n'$ representing the averaged population of the particle momentum distribution and a cutoff $\Lambda$ adjusted in such a way that particle number conservation is satisfied and by requiring the kinetic energy per nucleon to be  
%$35\, MeV$. 
\begin{equation}
 <t>=\frac{3}{k^3_F}\int_0^\Lambda dk k^2\,\frac{k^2}{2M}\,n_{\bf k}=35\,MeV,   
\end{equation}
the value discussed in section \ref{sec_emc} in the context of the EMC effect. 
For $\Delta n$ we take the value $0.15$, as discussed in section \ref{subsec_occupation_number}.
The result of this simplified calculation 
is  displayed on Fig.\ref{f1} where the correlated sum rule is compared with the Fermi gas one. 
The enhancement present at $q=0$ persists up to momentum $q=350\, MeV$.

%\newpage
\subsection{Fermi sea depopulation and kinetic energy per nucleon}
\label{subsec_explicit_calc_Dn_T}

As stressed in the introduction, one aim of this paper is to evaluate 
the compatibility of the approaches based on effective interactions (i.e., equivalent to a G matrix) with the expected amount of correlations 
from \textit{ab initio} calculations based on correlated wave functions. For this purpose we focus on the Fermi sea depopulation and on the kinetic energy per nucleon 
which are good indicators of the correlations increasing the response functions entering the neutrino-nucleus cross section. 
We consider first the approach of Ref. \cite{AEM84} which is the basis of our evaluation of the 2p-2h contribution to our neutrino-nucleus cross sections \cite{MECM2009,MECM2010,MEC2011,ME2013,ME2014}. 
It is also interesting to check whether the parameters used in our RPA calculations of the 1p-1h response lead for the Fermi sea depopulation and for the kinetic energy per nucleon to results compatible with the ones obtained by correlated waves functions approaches. 
In the same way we investigate the compatibility of chiral relativistic theories devoted to the study of the equations of state of nuclear matter or of neutron stars  \cite{CE2007,CE2011,MC2008,MC2009,MMC2013,Hu2009,Hu2010,Hu2011}. 
\subsubsection{Explicit calculations for $\pi+ \rho + g'$ interactions}
We list below the various formulas for the depopulation and the kinetic energy per nucleon, taking for the G matrix the $\pi+\rho+g'$  interaction. The occupation numbers entering the depopulation write: 
\begin{eqnarray}
\Delta n^h_{\bf k}&=&\frac{3}{2}\int\frac{d\bf t}{(2\pi)^3}\int\frac{4d{\bf p}_1}{(2\pi)^3}\int\frac{4d{\bf p}_2}{(2\pi)^3}\,(2\pi)^3\delta\left({\bf k}-{\bf p}_1-{\bf t}\right)\,\Theta\left(k_F-k\right)\,
\Theta\left(k_F-|{\bf p}_2 - {\bf t}|\right)
\nonumber\\
&&\left(\frac{g_A}{2 f_\pi}\right)^4 \bigg[\Theta\left(p_1-k_F\right)\,\Theta\left(p_2-k_F\right)
\left|\frac{v^2_{LNN}(t)+2\,v^2_{TNN}(t)}{\varepsilon_{{\bf p}_1}-\varepsilon_{{\bf p}_1+\bf t}+\varepsilon_{{\bf p}_2}-\varepsilon_{{\bf p}_2-\bf t}}\right|^2\nonumber\\
&&+\Theta\left(p_1-k_F\right)\,\frac{4}{9}R^2_{N\Delta}\left|\frac{v^2_{LN\Delta}(t)+2\,v^2_{TN\Delta}(t)}{\varepsilon_{{\bf p}_1}-\varepsilon_{{\bf p}_1+\bf t}+\varepsilon^\Delta_{{\bf p}_2}-\varepsilon_{{\bf p}_2-\bf t}}\right|^2\nonumber\\
&&+\Theta\left(p_2-k_F\right)\,\frac{4}{9}R^2_{N\Delta}\left|\frac{v^2_{LN\Delta}(t)+2\,v^2_{TN\Delta}(t)}{\varepsilon^\Delta_{{\bf p}_1}-\varepsilon_{{\bf p}_1+\bf t}+\varepsilon_{{\bf p}_2}-\varepsilon_{{\bf p}_2-\bf t}}\right|^2\nonumber\\
&&+\,\frac{16}{81}R^4_{N\Delta}\left|\frac{v^2_{L\Delta\Delta}(t)+2\,v^2_{T\Delta\Delta}(t)}{\varepsilon^\Delta_{{\bf p}_1}-\varepsilon_{{\bf p}_1+\bf t}+\varepsilon^\Delta_{{\bf p}_2}-\varepsilon_{{\bf p}_2-\bf t}}\right|^2\bigg]\\
\Delta n^N_{\bf k}&=&\frac{3}{2}\int\frac{d\bf t}{(2\pi)^3}\int\frac{4d{\bf h}_1}{(2\pi)^3}\int\frac{4d{\bf h}_2}{(2\pi)^3}\,(2\pi)^3\delta\left({\bf k}-{\bf h}_1-{\bf t}\right)\,\Theta\left(k-k_F\right)\,\nonumber\\
&&\Theta\left(k_F-h_1\right)
\,\Theta\left(k_F-h_2\right)\nonumber\\
&&\left(\frac{g_A}{2 f_\pi}\right)^4\bigg[\left|\frac{v^2_{LNN}(t)+2\,v^2_{TNN}(t)}{\varepsilon_{{\bf h}_1+\bf t}-\varepsilon_{{\bf h}_1}+\varepsilon_{{\bf h}_2-\bf t}-\varepsilon_{{\bf h}_2}}\right|^2\,\Theta\left(|{\bf h}_2 - {\bf t}|-k_F\right)\nonumber\\
&&+\frac{4}{9} R^2_{N\Delta}
\left|\frac{v^2_{LN\Delta}(t)+2\,v^2_{TN\Delta}(t)}{\varepsilon_{{\bf h}_1+\bf t}-\varepsilon_{{\bf h}_1}+\varepsilon^\Delta_{{\bf h}_2-\bf t}-\varepsilon_{{\bf h}_2}}\right|^2\bigg]\\
\Delta n^\Delta_{\bf k}&=&\frac{3}{2}\int\frac{d\bf t}{(2\pi)^3}\int\frac{4d{\bf h}_1}{(2\pi)^3}\int\frac{4d{\bf h}_2}{(2\pi)^3}\,(2\pi)^3\,\delta\left({\bf k}-{\bf h}_1-{\bf t}\right)\,\,\Theta\left(k_F-h_1\right)
\,\Theta\left(k_F-h_2\right)\left(\frac{g_A}{2 f_\pi}\right)^4\nonumber\\
&&\frac{4}{9}R^2_{N\Delta}\bigg[\left|\frac{v^2_{LN\Delta}(t)+2\,v^2_{TN\Delta}(t)}{\varepsilon^\Delta_{{\bf h}_1+\bf t}-\varepsilon_{{\bf h}_1}+\varepsilon_{{\bf h}_2-\bf t}-\varepsilon_{{\bf h}_2}}\right|^2\,\Theta\left(|{\bf h}_2 - {\bf t}|-k_F\right)\nonumber\\
&&+\,\frac{4}{9}R^2_{N\Delta}
\left|\frac{v^2_{L\Delta\Delta}(t)+2\,v^2_{T\Delta\Delta}(t)}{\varepsilon^\Delta_{{\bf h}_1+\bf t}-\varepsilon_{{\bf h}_1}+\varepsilon^\Delta_{{\bf h}_2-\bf t}-\varepsilon_{{\bf h}_2}}\right|^2\bigg],
\end{eqnarray}
where $g_A$ is the nucleon axial coupling constant, $f_\pi$ is the pion decay constant and $R_{N\Delta}=\frac{g_{\pi N\Delta}}{g_{\pi NN}}$. We have used the following notations:
\begin{eqnarray}
v_{LNN}(t)=v_\pi(t)+\Gamma^2_\pi(t)\, g'_{NN}\,\,\,  && \,\,\, v_{TNN}(t)=v_\rho(t)+\Gamma^2_\pi(t)\, g'_{NN} \nonumber\\
v_{LN\Delta}(t)=v_\pi(t)+\Gamma^2_\pi(t) \,g'_{N\Delta}\,\,\,   &&\,\,\,  v_{TN\Delta}(t)=v_\rho(t)+\Gamma^2_\pi(t)\, g'_{N\Delta} \nonumber\\
v_{L\Delta\Delta}(t)=v_\pi(t)+\Gamma^2_\pi(t)\, g'_{\Delta\Delta}\,\,\,   && \,\,\, v_{T\Delta\Delta}(t)=v_\rho(t)+\Gamma^2_\pi(t) \,g'_{\Delta\Delta}
\end{eqnarray}
\begin{eqnarray}
&&v_\pi(t)= -\Gamma^2_\pi(t)\frac{t^2}{t^2+m^2_\pi}\,\,\,v_\rho(t)= -\Gamma^2_\rho(t)\,C_\rho\frac{t^2}{t^2+m^2_\rho},%\nonumber\\
%&&\Gamma_\pi(t)=\left(\frac{\Lambda_\pi^2}{\Lambda_\pi^2+t^2}\right)^n, \Gamma_\rho(t)=\left(\frac{\Lambda_\rho^2}{\Lambda_\rho^2+t^2}\right)^n, R_{N\Delta}=\frac{g_{\pi N\Delta}}{g_{\pi NV}}=2 
\label{ffpirho}
\end{eqnarray}
where $g'_{NN}, g'_{N\Delta}, g'_{\Delta\Delta}$ are the Landau-Migdal parameters \cite{ISW06}, $C_\rho=1.5 \div 2.2$ is the usual factor entering the rho exchange interaction, $\Gamma_\pi(t)$ and $\Gamma_\rho(t)$ are the $\pi NN$ and $\rho NN$ form factors.\\
From baryon number conservation, the total depopulation per nucleon, $\Delta n$, is equal to the fraction of states (nucleon or $\Delta$) outside of the Fermi sea. It writes: 
\begin{eqnarray}
\Delta n &=&\Delta n^N \,+\,\Delta n^\Delta\\
\Delta n^N &=&\frac{1}{\rho}\,\int\frac{d\bf k}{(2\pi)^3}\Delta n^N_{\bf k}=
\frac{3}{2\rho}\int\frac{d\bf t}{(2\pi)^3}\int\frac{4d{\bf h}_1}{(2\pi)^3}\int\frac{4d{\bf h}_2}{(2\pi)^3}\,\Theta\left(|{\bf h}_1 + {\bf t}|-k_F\right)\,\,\Theta\left(k_F-h_1\right)\,\Theta\left(k_F-h_2\right)\nonumber\\&&\left(\frac{g_A}{2 f_\pi}\right)^4%\nonumber\\&&
\left[\left|\frac{v^2_{LNN}(t)+2\,v^2_{TNN}(t)}{\varepsilon_{{\bf h}_1+\bf t}-\varepsilon_{{\bf h}_1}+\varepsilon_{{\bf h}_2-\bf t}-\varepsilon_{{\bf h}_2}}\right|^2\,\Theta\left(|{\bf h}_2 - {\bf t}|-k_F\right)\,\right.
\nonumber\\&&+\,\left.\frac{4}{9}R^2_{N\Delta}
\left|\frac{v^2_{LN\Delta}(t)+2\,v^2_{TN\Delta}(t)}{\varepsilon_{{\bf h}_1+\bf t}-\varepsilon_{{\bf h}_1}+\varepsilon^\Delta_{{\bf h}_2-\bf t}-\varepsilon_{{\bf h}_2}}\right|^2\right]\\
\Delta n^\Delta &=&\frac{1}{\rho}\,\int\frac{d\bf k}{(2\pi)^3}\Delta n^\Delta_{\bf k}=
\frac{3}{2\rho}\int\frac{d\bf t}{(2\pi)^3}\int\frac{4d{\bf h}_1}{(2\pi)^3}\int\frac{4d{\bf h}_2}{(2\pi)^3}\,\Theta\left(k_F-h_1\right)\,\Theta\left(k_F-h_2\right)\nonumber\\
&&\left(\frac{g_A}{2 f_\pi}\right)^4\,\frac{4}{9}R^2_{N\Delta}%\nonumber\\&&
\left[\left|\frac{v^2_{LN\Delta}(t)+2\,v^2_{TN\Delta}(t)}{\varepsilon^\Delta_{{\bf h}_1+\bf t}-\varepsilon_{{\bf h}_1}+\varepsilon_{{\bf h}_2-\bf t}-\varepsilon_{{\bf h}_2}}\right|^2\,\Theta\left(|{\bf h}_2 - {\bf t}|-k_F\right)\,\right.\nonumber\\
&&
\left.+\,\frac{4}{9}R^2_{N\Delta}
\left|\frac{v^2_{L\Delta\Delta}(t)+2\,v^2_{T\Delta\Delta}(t)}{\varepsilon^\Delta_{{\bf h}_1+\bf t}-\varepsilon_{{\bf h}_1}+\varepsilon^\Delta_{{\bf h}_2-\bf t}-\varepsilon_{{\bf h}_2}}\right|^2\right].
\label{depop}
\end{eqnarray}
The kinetic energy per nucleon can be decomposed into: 
\begin{equation}
<t>=\frac{3}{5}\frac{k^2_F}{2M}\,-\,<t>^{depop}\,+\,<t>_{tail}^{N}\,+\,<t>^{\Delta}.
\end{equation}
The explicit form for each of these components is, after appropriate exchange of integration variable :
\begin{eqnarray}
<t>^{depop}&=&\frac{1}{\rho}\,\int\frac{d\bf k}{(2\pi)^3}\frac{k^2}{2M}\,\Theta\left(k_F-k\right)\,\Delta n^h_{\bf k}\nonumber\\
&=&\frac{3}{2\rho}\int\frac{d\bf t}{(2\pi)^3}\int\frac{4d{\bf p}_1}{(2\pi)^3}\int\frac{4d{\bf p}_2}{(2\pi)^3}\,
\frac{|{\bf p}_1 + {\bf t}|^2}{2M}\,\Theta\left(k_F-|{\bf p}_1 + {\bf t}|\right)
\Theta\left(k_F-|{\bf p}_2 - {\bf t}|\right)\,\nonumber\\&&\left(\frac{g_A}{2 f_\pi}\right)^4
%\nonumber\\&&
\bigg[\Theta\left(p_1-k_F\right)\,\Theta\left(p_2-k_F\right)
\left|\frac{v^2_{LNN}(t)+2\,v^2_{TNN}(t)}{\varepsilon_{{\bf p}_1}-\varepsilon_{{\bf p}_1+\bf t}+\varepsilon_{{\bf p}_2}-\varepsilon_{{\bf p}_2-\bf t}}\right|^2\nonumber\\
&&+\Theta\left(p_1-k_F\right)\,\frac{4}{9}R^2_{N\Delta}\left|\frac{v^2_{LN\Delta}(t)+2\,v^2_{TN\Delta}(t)}{\varepsilon_{{\bf p}_1}-\varepsilon_{{\bf p}_1+\bf t}+\varepsilon^\Delta_{{\bf p}_2}-\varepsilon_{{\bf p}_2-\bf t}}\right|^2\nonumber\\
&&
+\Theta\left(p_2-k_F\right)\,\frac{4}{9}R^2_{N\Delta}\left|\frac{v^2_{LN\Delta}(t)+2\,v^2_{TN\Delta}(t)}{\varepsilon^\Delta_{{\bf p}_1}-\varepsilon_{{\bf p}_1+\bf t}+\varepsilon_{{\bf p}_2}-\varepsilon_{{\bf p}_2-\bf t}}\right|^2\nonumber\\
&&+\,\frac{16}{81}R^4_{N\Delta}\left|\frac{v^2_{L\Delta\Delta}(t)+2\,v^2_{T\Delta\Delta}(t)}{\varepsilon^\Delta_{{\bf p}_1}-\varepsilon_{{\bf p}_1+\bf t}+\varepsilon^\Delta_{{\bf p}_2}-\varepsilon_{{\bf p}_2-\bf t}}\right|^2\bigg]\\
<t>_{tail}^{N}&=&\frac{1}{\rho}\,\int\frac{d\bf k}{(2\pi)^3}\frac{k^2}{2M}\,\Theta\left(k-k_F\right)\Delta n^N_{\bf k}\nonumber\\
&=&\frac{3}{2\rho}\int\frac{d\bf t}{(2\pi)^3}\int\frac{4d{\bf h}_1}{(2\pi)^3}\int\frac{4d{\bf h}_2}{(2\pi)^3}\,
\frac{|{\bf h}_1 + {\bf t}|^2}{2M}\Theta\left(|{\bf h}_1 + {\bf t}|-k_F\right)\,\,\Theta\left(k_F-h_1\right)\,\Theta\left(k_F-h_2\right)\nonumber\\&&\left(\frac{g_A}{2 f_\pi}\right)^4%\nonumber\\&&
\left[\left|\frac{v^2_{LNN}(t)+2\,v^2_{TNN}(t)}{\varepsilon_{{\bf h}_1+\bf t}-\varepsilon_{{\bf h}_1}+\varepsilon_{{\bf h}_2-\bf t}-\varepsilon_{{\bf h}_2}}\right|^2\,\Theta\left(|{\bf h}_2 - {\bf t}|-k_F\right)\,\right.
\nonumber\\&&\left.+\,\frac{4}{9}R^2_{N\Delta}
\left|\frac{v^2_{LN\Delta}(t)+2\,v^2_{TN\Delta}(t)}{\varepsilon_{{\bf h}_1+\bf t}-\varepsilon_{{\bf h}_1}+\varepsilon^\Delta_{{\bf h}_2-\bf t}-\varepsilon_{{\bf h}_2}}\right|^2\right]\\
<t>^\Delta&=&\frac{1}{\rho}\,\int\frac{d\bf k}{(2\pi)^3}\left(\omega_\Delta +\frac{k^2}{2 M_\Delta}\right)\Delta n^\Delta_{\bf k}\nonumber\\
&=&\frac{3}{2\rho}\int\frac{d\bf t}{(2\pi)^3}\int\frac{4d{\bf h}_1}{(2\pi)^3}\int\frac{4d{\bf h}_2}{(2\pi)^3}\,\left(\omega_\Delta +\frac{|{\bf h}_1 + {\bf t}|^2}{2 M_\Delta}\right)\Theta\left(k_F-h_1\right)\,\Theta\left(k_F-h_2\right)\nonumber\\&&\left(\frac{g_A}{2 f_\pi}\right)^4\,\frac{4}{9}R^2_{N\Delta}
%\nonumber\\&&
\left[\left|\frac{v^2_{LN\Delta}(t)+2\,v^2_{TN\Delta}(t)}{\varepsilon^\Delta_{{\bf h}_1+\bf t}-\varepsilon_{{\bf h}_1}+\varepsilon_{{\bf h}_2-\bf t}-\varepsilon_{{\bf h}_2}}\right|^2\,\Theta\left(|{\bf h}_2 - {\bf t}|-k_F\right)\,\right.
\nonumber\\&&\left.
+\,\frac{4}{9}R^2_{N\Delta}
\left|\frac{v^2_{L\Delta\Delta}(t)+2\,v^2_{T\Delta\Delta}(t)}{\varepsilon^\Delta_{{\bf h}_1+\bf t}-\varepsilon_{{\bf h}_1}+\varepsilon^\Delta_{{\bf h}_2-\bf t}-\varepsilon_{{\bf h}_2}}\right|^2\right] .
\end{eqnarray}
These formulas are used in the following  numerical evaluations for different values of the interaction parameters. 

\subsubsection{Numerical results}
Our  numerical estimates are made for a density $\rho=0.8 \rho_0$, i.e., $k_F=245\,MeV$, appropriate for the carbon nucleus.  For this nucleus correlated wave functions calculations give, as mentioned in sections \ref{sec_emc}, \ref{subsec_occupation_number}, \ref{subsection_sum_rule}, $\Delta n \simeq 0.15\div0.2$ and $<t>\simeq35~MeV$. We take these values to test our effective approaches. \\

\begin{figure}
 \centering
  \includegraphics[width=0.8\textwidth]{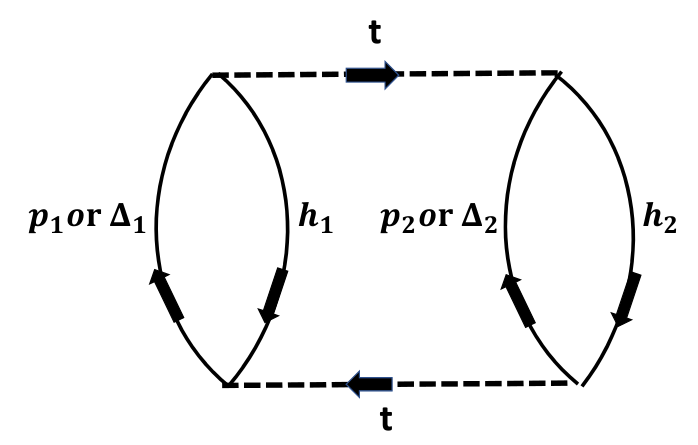}
  \caption{Diagrammatic view of the mechanisms contributing to the population of particle state beyond the Fermi energy. Replacing $p_1$ by $\Delta_1$ in the left bubble corresponds to the presence of the Delta resonance in the ground state.}
  \label{f2}
\end{figure}

In the approach of Ref. \cite{AEM84}, a unique Landau-Migdal parameter,
$g'_{NN}= g'_{N\Delta}= g'_{\Delta\Delta}=0.5,$ 
and a monopole $\pi NN$ form factor with a cutoff $\Lambda_\pi=0.9\,GeV$ were used. For the calculation of the 2p-2h contributions to the response function, rho exchange was ignored so as to avoid too much suppression of the tensor force. With these parameters and ignoring the presence of the $\Delta$, we find for the depopulation of the Fermi sea $\Delta n^N=0.149$. If we add the influence of the Delta ($\Delta_2$ in the right bubble of Fig. \ref{f2}), the result becomes $\Delta n^N=0.149+0.032=0.181$. The direct contribution of Delta states in the ground state ($\Delta_1$ in the left bubble) is $\Delta n^\Delta=0.032+0.028=0.06$, the second number corresponding to the presence of the Delta in the right bubble of Fig. \ref{f2}. Grouping everything together, one finds : 
$$\Delta n^{depop}=\Delta n^N  + \Delta n^\Delta =0.24,$$
quite a large number, with $6\,\%$ of Deltas in the ground state. For the kinetic energy per nucleon, we find:
\begin{eqnarray}
<t>&=&<t>^{FG}\,-\,<t>^{depop}\,+\,<t>_{tail}^{N}\,+\,<t>^{\Delta}\nonumber\\
&=& 19.0\,-4.8\,+\,18.7\,+\,18.5\,= 51.4\, MeV.
\end{eqnarray}
There is a considerable contribution of the Delta, presumably due to the conjunction of two factors: 
on the one hand the hard form factor (which leads an unphysically large contribution at high momenta, in particular in the transverse channel, where rho exchange is ignored); on the other hand the use of non relativistic kinematics which overestimates high momenta contributions. 

In order to make a comparison with the work of Ref. \cite{AEM84}, we notice that the Deltas are not incorporated in the right bubble. If we omit these $\Delta_2$ contributions in our calculation we find:
$$\Delta n^{depop}=\Delta n^N  + \Delta n^\Delta =0.149\,+\,0.032\,=0.181. $$
with now only $3\,\%$ of Deltas in the ground state. The kinetic energy per nucleon is reduced as well:
\begin{eqnarray}
<t>&=&<t>^{FG}\,-\,<t>^{depop}\,+\,<t>_{tail}^{N}\,+\,<t>^{\Delta}\nonumber\\
&=& 19.0\,-3.6\,+\,13.2\,+\,9.8\,= 38.4\, MeV
\end{eqnarray}
which is not far from the values obtained from microscopic calculations with correlated wave functions. 

Since the results of Ref. \cite{AEM84} were used as inputs
in our evaluations \cite{MECM2009} of the 2p-2h contributions to the neutrino cross-section, we can conclude that our parameters were consistent with the amount of short range correlations found in the correlated wave functions approaches.\\

The question is if effective interactions which incorporate rho meson exchange are also compatible with correlated wave functions results. Such interactions were used in the RPA calculations 
of neutrino-nucleus cross sections by the Lyon \cite{MECM2009} 
and Valencia \cite{Nieves2011} groups, as well as in the nuclear matter calculations of the Osaka group \cite{Hu2010}. 
Typical parameter-set used in the p-h interaction (that we identify with the underlying G matrix) is: 
$$g'_{NN}=0.7,\,\, g'_{N\Delta}= g'_{\Delta\Delta}=0.5,\,\,\, C_\rho=2, \,\,\,\,R_{N\Delta}=\frac{g_{\pi N\Delta}}{g_{\pi NN}}=2.$$ 
We take the following form factors :
\begin{equation}
\Gamma_\pi(t)=\frac{\Lambda_\pi^2}{\Lambda_\pi^2+t^2},\quad \Gamma_\rho(t)=\frac{\Lambda_\rho^2 - m^2_\rho}{\Lambda_\rho^2+t^2},\quad\hbox{with}  \,\,\Lambda_\pi=1\,GeV,\quad\Lambda_\rho=1.3\,GeV.
\end{equation}
With these inputs we find:
\begin{eqnarray}
\Delta n^N=0.151 + 0.018,%\qquad
%\hbox{the 2nd number refers to a $\Delta$ in the right bubble of Fig.4}
\nonumber\\
\Delta n^\Delta=0.018+0.017,
%\qquad \hbox{the 2nd number refers to a $\Delta$ in the right bubble of Fig.4}
\nonumber
\end{eqnarray}
where the second number refers to a $\Delta$ in the right bubble of Fig.\ref{f2}. 
This gives for the Fermi sea depopulation: 
$$\Delta n^{depop}=\Delta n^N  + \Delta n^\Delta =0.20,$$
and $3.5\,\%$ of Deltas in the ground state.\\ 
For the kinetic energy per nucleon we obtain :
\begin{eqnarray}
<t>&=&<t>^{FG}\,-\,<t>^{depop}\,+\,<t>_{tail}^{N}\,+\,<t>^{\Delta}\nonumber\\
&=& 19.0\,-4.2\,+\,15.1\,+\,10.8\,= 40.7\, MeV. \label{tcut}
\end{eqnarray}

From these two examples we observe that the results are  sensitive to the parameters affecting particularly the high momentum behaviour of the transverse interaction. In order to check the convergence of the momentum $t$ integration, we cut the integrals at $t=\Lambda_\rho=1.3\,GeV$. 
The occupation numbers 
are slightly affected through the Delta contributions which are decreased by about $10\,\%$. For the depopulation we find: $\Delta n^{depop}= 0.197$ in place of $0.204$. The kinetic energy instead is more affected: 
\begin{eqnarray}
<t>&=&<t>^{FG}\,-\,<t>^{depop}\,+\,<t>_{tail}^{N}\,+\,<t>^{\Delta}\nonumber\\
&=& 19.02\,-4.05\,+\,12.6\,+\,9.3\,= 36.9\, MeV,
\end{eqnarray}
instead of $<t>=40.7 MeV$. 

Our conclusions on the compatibility of effective approaches with microscopic correlated wave functions calculations are nevertheless not affected. 
It would be relevant to incorporate the relativistic kinematics which may act as a cut in the calculation of the various components of the kinetic energy.\\

As mentioned before it is interesting to check as well whether effective theories devoted to the study of the equation of state and chiral properties (chiral condensate and QCD susceptibilities) of nuclear matter are compatible with the expected amount of correlations. In Ref. \cite{CE2007}, we used: 
$$g'_{NN}=0.7,\,\, g'_{N\Delta}=0.3,\,\, g'_{\Delta\Delta}=0.5,\,\,\, C_\rho=2, \,\,\,\,R_{N\Delta}=\frac{g_{\pi N\Delta}}{g_{\pi NN}}=\sqrt{\frac{72}{25}}.$$
At variance with the previous cases, the dipole $\pi NN$ form factor:
$$
\Gamma_\pi(t)=\left(\frac{\Lambda_\pi^2}{\Lambda_\pi^2+t^2}\right)^2,\,\,\hbox{with}\,\,\,\Lambda_\pi=0.98\,GeV
$$
was selected from the lattice calculations of Ref. \cite{LTY2004} for the evolution of the nucleon mass with the light quark mass, driven by the nucleon sigma term.  
More precisely, the value of $\Lambda_\pi=0.98\,GeV$ was fixed in order to obtain a pion cloud contribution to the nucleon sigma term of $\sigma^{(\pi)}_N=21.5\,MeV$ \cite{CE2007}.
%This approach used the Finite Range Regularization (FFR) method for the chiral extrapolation of the nucleon mass towards its physical value. More precisely the a pion cloud contribution to the nucleon sigma term of $\sigma^{(\pi)}_N=21.5\,MeV$ \cite{CE2007}. 
We also used the same dipole form factor for the rho exchange in order to limit the contribution of high momenta. 
With these choices the convergence properties of all the $t$ integrals are excellent % and we can be confident in the high momentum behaviour of the $t$ integrals since it is actually constrained by nucleon physical properties, namely the sigma commutator.
and we find for the contributions to the Fermi sea depopulation:
\begin{eqnarray}
\Delta n^N=0.07 + 0.01,%\qquad\hbox{the second number corresponds to a Delta in the right bubble}
\nonumber\\
\Delta n^\Delta=0.01+0.003,%\qquad\hbox{the second number corresponds to a Delta in the right bubble}
\nonumber
\end{eqnarray}
where the second number refers to a $\Delta$ in the right bubble of Fig.\ref{f2}. This gives for the total Fermi sea depopulation: 
$$\Delta n^{depop}=\Delta n^N  + \Delta n^\Delta =0.09.$$
with only  $1.1\,\%$ of Deltas in the ground state.\\ 
As for the kinetic energy per nucleon we obtain :
\begin{eqnarray}
<t>&=& <t>^{FG}\,-\,<t>^{depop}\,+\,<t>_{tail}^{N}\,+\,<t>^{\Delta}\nonumber\\
&=& 19.0\,-\,1.9\,+\,5.4\,+\,3.5\,= 26.0\, MeV.
\end{eqnarray}
We find that the occupation number and the kinetic energy are significantly reduced, with a moderate role of the Delta states. 
However the internal consistency of the approach 
would require to incorporate also $\sigma +\omega$ exchange on top the $\pi+\rho$ exchange. With the same form factors as previously, we find an additional contribution $\Delta n^{\sigma +\omega}=0.074$ and  $<t>^{\sigma +\omega}=-1.505 +9.97=8.26\,MeV$. Adding all the exchange contributions we finally obtain: 
\begin{eqnarray}
\hbox{With Delta}:\,\Delta n^{depop}=0.165, \,\,<t>=34.2 MeV\nonumber\\
\hbox{Without Delta}:\,\Delta n^{depop}=0.149, \,\,<t>=30.6 MeV,\nonumber
\end{eqnarray}
which is compatible with microscopic calculations based on correlated wave functions.\\

%\GC{old version The first thing to be checked is whether this approach yields a correct value for the Fermi sea depopulation and for the kinetic energy per nucleon.} \\

Although these numerical results are interesting and encouraging, they should be considered only as a starting point for two main reasons: the remaining uncertainties related to the modeling of the effective interaction (\textit{i.e.} $G$ matrix) and the formal improvement of the underlying one-body Green's function with its full energy dependence which may affect the momentum distribution.

\section{Conclusions}

We have investigated the compatibility of two different approaches for the treatment of the nuclear correlations. 
On the one hand the \textit{``ab-initio''} calculations based on correlated wave functions starting from bare nucleon-nucleon interaction. On the other hand independent particle models (such as Hartree-Fock calculations) using phenomenological interactions. 
After a formal analysis on the correspondence between the two approaches, we evaluated whether approaches based on effective interactions are compatible with the expected amount of correlations coming from \textit{ab initio} calculations. 
Focusing on the Fermi sea depopulation and on the kinetic energy per nucleon, we have found that this is indeed the case. 
As a next step we plan to extend the present analysis to the nuclear responses and to the neutrino-nucleus cross sections.

\newpage
\appendix 

%%%%%%%%%%%%%%%%%%%%%%%%%%%%%%%%%%%%%%%%%%%%%%%%
\section{The one-body Green's function and the mass operator}
%%%%%%%%%%%%%%%%%%%%%%%%%%%%%%%%%%%%%%%%%%%%%%%
\label{appendix_G_and_M}
In this appendix we remind some textbook formulas (see for example Ref. \cite{fetter_walecka}) of the many-body theory.\\
The  one-body Green's function is defined as : 
\begin{equation}
G_{k',k}(t)=<0|-i\,{\cal{T}}\left(a_{k'}(t) , a^\dagger_{k}(0) \right)|0>=\int \frac{d\Omega}{2\pi} e^{-i\Omega t} G_{k',k}(\Omega). 
\end{equation}
Its Lehmann representation can be written in a dispersive form as :
\begin{equation}
\label{eq_lehmann}
 G_{k',k}(\Omega) =\int dE\left(\frac{S^p_{k',k}(E)}{\Omega-E+i\eta}+\frac{S^h_{k,k'}(E)}{\Omega-E-i\eta}\right)
\end{equation}
and is enterely known once the hole and particle spectral functions, defined as 
\begin{eqnarray}
S^h_{k,k'}(E)&=&\sum_n <0|a^\dagger_k|n>< n|a_{k'}|0>\,
\delta\left(E+E_n^{A-1}-E_0^A\right)  \nonumber\\ 
S^p_{k',k}(E)&=&\sum_n <0|a_{k'}|n>< n|a^\dagger_k|0>\,
\delta\left(E-E_n^{A+1}+E_0^A\right),  
\end{eqnarray}
are known.\\
The single particle propagator in nuclear matter is obtained as the diagonal component one-body Green's function:
\begin{equation}
G(E,{\bf k})= \int dt\, e^{iEt}\left\langle 0\right|-iT\left(a_{\bf k}(t), a^\dagger_{\bf k}(0)\right)\left|0\right\rangle=\left(E-(k^2/2M)-M(E,{\bf k})\right)^{-1}, %\nonumber\\
\end{equation}
where $M(E,{\bf k})$ is the mass operator.\\ 
The real and imaginary part of the nuclear matter single particle propagator can be expressed in terms of the spectral functions as: 
\begin{eqnarray}
Re\, G(E,{\bf k})&=&\int dE'\,\frac{S^p(E',{\bf k})+S^h(E',{\bf k})}{E-E'}\\
Im\, G(E,{\bf k})&=& -\pi\left(S^p(E,{\bf k})-S^h(E,{\bf k})\right).
\end{eqnarray}
The imaginary part of the mass operator $M(E,{\bf k})$ (i.e. the optical potential), assuming  that it is always smaller than the real part (see Ref. \cite{Schuck89})%(see section 3.3 of \cite{Schuck89})
, is related to the single particle propagator and to the spectral functions by the following relations :
\begin{equation}
\label{eq_appendix_im_M}
-\frac{1}{\pi} Im\, G(E,{\bf k})\simeq -\frac{1}{\pi}\frac{Im\, M(E,{\bf k})}{(E-\epsilon_k)^2} =S^p(E,{\bf k})-S^h(E,{\bf k}),
\end{equation}
where $\epsilon_k$ is the Hartree-Fock single particle energy. Inserting in Eq. (\ref{eq_appendix_im_M}) the expression of the spectral functions given by Eqs. (\ref{Spectral2}),(\ref{Spectral3}), we obtain the expression of Eq.(\ref{Mass}) for the imaginary part of the mass operator.

%%%%%%%%%%%%%%%%%%%%%%%%%%%%%%%%%%%%%%%%%%%%%%%%%
\section{Response function in the factorization approximation}
\label{appendix_factorization}
%%%%%%%%%%%%%%%%%%%%%%%%%%%%%%%%%%%%%%%%%%%%%%%%%
In this appendix we explicitly derive the expression of the response function in the factorization approximation. The response function, already defined in Eq.(\ref{eq_def_resp}), can be explicitly written as: 
\begin{eqnarray}
R(\omega,{\bf q})&=&\sum_{n}\,|< n|\sum_{j=1}^A\,{\cal{O}}(j)\, e^{i{\bf q}\cdot{\bf x}_j} |0 >|^2\,\delta(\omega-E_n + E_0)\nonumber\\
&=&\sum_{n}\sum_{k'_1, k'_2,k_1,k_2} <k'_1|{\cal{O}}^\dagger  e^{-i{\bf q}\cdot{\bf x}}|k'_2><k_2|{\cal{O}} e^{i{\bf q}\cdot{\bf x}}|k_1>
<0|a^\dagger_{k'_1}a_{k'_2}|n><n|a^\dagger_{k_2} a_{k_1}|0>\nonumber\\
&\equiv&\left(-\frac{1}{\pi}\right)  Im \Pi(\omega,{\bf q}).
\end{eqnarray}
The response function is the imaginary part of the polarization propagator $\Pi(\omega,{\bf q})$, which can be expressed as:
\begin{eqnarray}
\label{eq_appendix_Pi_factorized}
\Pi(\omega,{\bf q})&=&\int dt\, e^{i\omega t}\,\sum_{k'_1, k'_2,k_1,k_2} <k'_1|{\cal{O}}^\dagger  e^{-i{\bf q}\cdot{\bf x}}|k'_2><k_2|{\cal{O}} e^{i{\bf q}\cdot{\bf x}}|k_1> \,\nonumber\\
&&
<0|-i\,{\cal{T}}\left(a^\dagger_{k'_1}(t)a_{k'_2}(t) , a^\dagger_{k_2}(0) a_{k_1}(0)\right)|0> \nonumber \\
&\approx&\int idt\, e^{i\omega t}\,\sum_{k'_1, k'_2,k_1,k_2} <k'_1|{\cal{O}}^\dagger  e^{-i{\bf q}\cdot{\bf x}}|k'_2><k_2|{\cal{O}} e^{i{\bf q}\cdot{\bf x}}|k_1> <0|-i\,{\cal{T}}\left(a^\dagger_{k'_1}(t) ,  a_{k_1}(0)\right)|0>\nonumber\\
&&<0|-i\,{\cal{T}}\left(a_{k'_2}(t) , a^\dagger_{k_2}(0) \right)|0>\nonumber\\
&=&\int idt\, e^{i\omega t}\,\sum_{k'_1, k'_2,k_1,k_2} <k'_1|{\cal{O}}^\dagger  e^{-i{\bf q}\cdot{\bf x}}|k'_2><k_2|{\cal{O}} e^{i{\bf q}\cdot{\bf x}}|k_1> (-1) G_{k_1,k'_1}(-t) G_{k'_2,k_2}(t),\nonumber\\
\end{eqnarray}
where we have applied the factorization scheme which consists to approximate the two-body Green's function as a product of two one-body Green's functions. 
Using the Lehmann representation, given in Eq.(\ref{eq_lehmann}),  for the one-body Green's function, one obtains for the polarization propagator : 
\begin{eqnarray}
\label{eq_Pi_Sh_Sp}
\Pi(\omega,{\bf q})&=&\sum_{k'_1, k'_2,k_1,k_2} <k'_1|{\cal{O}}^\dagger  e^{-i{\bf q}\cdot{\bf x}}|k'_2><k_2|{\cal{O}} e^{i{\bf q}\cdot{\bf x}}|k_1>\nonumber\\
&&\times(-i)\int \frac{d\Omega_1}{2\pi}\frac{d\Omega_2}{2\pi} \,2\pi\,\delta(\omega-\Omega_2+\Omega_1)\,
G_{k_1,k'_1}(\Omega_1) G_{k'_2,k_2}(\Omega_2)\nonumber\\
&=&\sum_{k'_1, k'_2,k_1,k_2} <k'_1|{\cal{O}}^\dagger  e^{-i{\bf q}\cdot{\bf x}}|k'_2><k_2|{\cal{O}} e^{i{\bf q}\cdot{\bf x}}|k_1>\nonumber\\
&&\times\int\frac{dE_1}{2\pi}\frac{dE_2}{2\pi}\left(\frac{S^h_{k_1,k'_1}(E_1)S^p_{k'_2,k_2}(E_2)}{\omega -E_2+E_1+ i\eta}
-\frac{S^p_{k_1,k'_1}(E_1)S^h_{k'_2,k_2}(E_2)}{\omega -E_2+E_1- i\eta}\right).
\end{eqnarray}
The response function is obtained by taking the imaginary part (for positive $\omega$) of Eq.(\ref{eq_Pi_Sh_Sp}), leading to:
\begin{eqnarray}
R(\omega,{\bf q})&=&\sum_{k'_1, k'_2,k_1,k_2} <k'_1|{\cal{O}}^\dagger  e^{-i{\bf q}\cdot{\bf x}}|k'_2><k_2|{\cal{O}} e^{i{\bf q}\cdot{\bf x}}|k_1>\nonumber\\
&&\int_{-\infty}^{\varepsilon_F} dE_1 \int_{\varepsilon_F}^{\infty} dE_2\, S^h_{k_1,k'_1}(E_1)\,S^p_{k'_2,k_2}(E_2)\,\delta(\omega-E_2+E_1).
\end{eqnarray}

\end{document}